\documentclass[useAMS,graphicx,usenatbib]{mn2e}

\usepackage{graphicx}
\usepackage{amssymb}

\def\lsim{\mathrel{\rlap{\lower4pt\hbox{\hskip1pt$\sim$}}
    \raise1pt\hbox{$<$}}}                
\def\gsim{\mathrel{\rlap{\lower4pt\hbox{\hskip1pt$\sim$}}
    \raise1pt\hbox{$>$}}}                

\title[A deep i-selected galaxy catalogue]{A deep i-selected
  multi-waveband galaxy catalogue in the COSMOS field\footnotemark[2]}

\author[Armin Gabasch et al.]{A. Gabasch$^{1,2,3}$\footnotemark[1], Y.
  Goranova$^{1,2,5}$, U. Hopp$^{1,2}$, S. Noll$^{1,4}$ and M. Pannella$^1$\vspace*{.25em}\\
  $^{1}$Max-Planck-Institut f\"ur extraterrestrische Physik,
  Giessenbachstr.,
  Postfach 1312, D-85741 Garching, Germany\\
  $^{2}$Universit\"atssternwarte M\"unchen, Scheinerstr. 1, D-81673
  M\"unchen, Germany\\
  $^{3}$European Southern Observatory, 85748 Garching, Germany\\
  $^{4}$Observatoire Astronomique Marseille Provence, Laboratoire
  d'Astrophysique de Marseille, Traverse du Siphon, 13376 Marseille
  Cedex 12, France\\
  $^5$Leiden Observatory, P.O. Box 9513, NL--2300 RA Leiden, The
  Netherlands }

\date{Accepted .... Received ....; in original form ....}

\pagerange{\pageref{firstpage}--\pageref{lastpage}} \pubyear{2007}

\begin{document}

\label{firstpage}

\maketitle

\begin{abstract}
  
  In this paper we present a deep and homogeneous i-band selected
  multi-waveband catalogue in the COSMOS field covering an area of
  about $0.7\sq\degr$. Our catalogue with a formal 50\% completeness
  limit for point sources of $i\sim 26.7$ comprises about 290~000
  galaxies with information in 8 passbands. We combine publicly
  available u, B, V, r, i, z, and K data with proprietary imaging in H
  band. We discuss in detail the observations, the data reduction, and
  the photometric properties of the H-band data. We estimate
  photometric redshifts for all the galaxies in the catalogue. A
  comparison with 162 spectroscopic redshifts in the redshift range $
  0 \lsim z \lsim 3$ shows that the achieved accuracy of the
  photometric redshifts is \mbox{$\Delta z / (z_{spec}+1) \lsim
    0.035$} with only $\sim 2$\% outliers. We derive absolute UV
  magnitudes and investigate the evolution of the luminosity function
  evaluated in the restframe UV (1500~\AA).  { There is a good
    agreement between the LFs derived here and the LFs derived in the
    FORS Deep Field. We see a similar brightening of M$^\ast$ and a
    decrease of $\phi^\ast$ with redshift.}
  The catalogue including the photometric redshift information is made
  publicly available.
\end{abstract}


\footnotetext[1]{E-mail: agabasch@eso.org}

\footnotetext[2]{Based on observations collected at the Centro
  Astron\'omico Hispano Alem\'an (CAHA), operated jointly by the
  Max-Planck-Institut f\"ur Astronomie, Heidelberg, and the Instituto
  de Astrofisica de Andalucia (CSIC).}

\section{Introduction}
\label{sec:intro}

In the last decade our knowledge about the evolution of global galaxy
properties over a large redshift range has improved considerably.  The
2dF Galaxy Redshift Survey (2dFGRS; \citealt{colless:1}), the Sloan
Digital Sky Survey (SDSS; \citealt{stoughton:1}), and the 2MASS survey
\citep{2MASS} have provided very large local galaxy samples with
spectroscopic and/or photometric information in various passbands.  Thanks
to these data sets we are now able to assess very accurate local
($z\sim 0.1$) reference points for many galaxy evolution measurements
like the luminosity function, the star formation activity, the spatial
clustering of galaxies, the stellar population, the morphology, etc.

In the redshift range between $0.2 \lsim z \lsim 1 $ pioneering work
has been done in the context of the Canada France Redshift Survey
\citep{lilly:3}, the Autofib survey \citep{ellis:1} and in the
Canadian Network for Observational Cosmology survey \citep{yee:1}.
They provide accurate distances and absolute luminosities by
spectroscopic followup of optically selected galaxies, thus being able
to probe basic properties of galaxy evolution. Moreover the
K20-survey \citep{cimatti:2} as well as the MUNICS survey
\citep{drory:2,feulner:1} extend the analysis into the near infrared
regime (for $0.2 \lsim z \lsim 1.5 $).

An important step towards probing the galaxy properties also in the
high redshift regime around $z \sim 3$ and $z\sim 4$ was the work of
\citet{steidel_lbg:1} and \citet{steidel_lbg:2}. They used colour
selection to discriminate between low and high redshift galaxies
\citep[see ][for a review]{giavalisco:3}. The so-called Lyman-break
galaxies (LBGs, mainly starburst galaxies at high redshift) are
selected by means of important features in the UV spectrum of
star-forming galaxies.

The next milestones in pushing the limiting magnitude for detectable
galaxies to fainter and fainter limits were the space based Hubble
Deep Field North (HDFN; \citealt{HDF96}) and Hubble Deep Field South
\citep[HDFS; ][]{HDFS00,HDFS00a} \citep[see ][for a
review]{ferguson:1}. Although of a limited field of view of about
$5\sq\arcmin$ only, the depth of the HDFs allowed the detection of
galaxies up to a redshift of 5 and even beyond.

In the past years the space based HDFs were supplemented by many more
multi-band photometric surveys like the NTT SUSI deep Field (NDF;
\citealt{arnouts_ntt}), the Chandra Deep Field South (CDFS;
\citealt{arnouts_cdfs}), the William Herschel Deep Field (WHDF;
\citealt{mccracken:1,metcalf:1}), the Subaru Deep Field/Survey (SDF;
\citealt{maihara:1,ouchi:2}), the COMBO-17 survey \citep{combo17:1},
FIRES \citep{labbe:1}, the FORS Deep Field (FDF; \citealt{fdf_data}),
the Great Observatories Origins Deep Survey (GOODS;
\citealt{giavalisco:1}), the Ultra Deep Field (UDF and UDF-Parallel
ACS fields; \citealt{giavalisco:2,bunker:1, bouwens:2004}), the VIRMOS
deep survey \citep{lefevre:3}, GEMS \citep{rix:1}, the Keck Deep
Fields \citep{sawicki:2}, and the Multiwavelength Survey by Yale-Chile
(MUSYC; \citealt{gawiser:1, quadri:1}).

With the advent of all these deep multi-band photometric surveys the
photometric redshift technique (essentially a generalisation of the
drop-out technique) can be used to identify high-redshift galaxies.
Photometric redshifts are often determined by means of template
matching algorithm that applies Bayesian statistics and uses
semi-empirical template spectra matched to broad-band photometry (see
also \citealt{baum:1, koo:1, brunner:1, Soto:1, benitez:1, bender:1,
  borgne:1, firth:1}).  Redshifts of galaxies that are several
magnitudes fainter than typical spectroscopic limits can be determined
reliably with an accuracy of \mbox{$\Delta z / (z_{spec}+1)$} of 0.02
to 0.1.

In this context the COSMOS survey (\citet{scoville:1}; see also
http://www.astro.caltech.edu/cosmos/ for an overview) combines deep to
very-deep multi-waveband information in order to extend the analysis
of deep pencil beam surveys to a much bigger volume, thus being able
to drastically increase the statistics and detect also very rare
objects. For this, the survey covers an area of about $2\sq\degr$ with
imaging by space-based telescopes (Hubble, Spitzer, GALEX, XMM,
Chandra) as well as large ground based telescopes (Subaru, VLA,
ESO-VLT, UKIRT, NOAO, CFHT, and others).

In this paper we combine publicly available u, B, V, r, i, z, and K
COSMOS data with proprietary imaging in the H band to derive a
homogeneous multi-waveband catalogue suitable for deriving accurate
photometric redshifts. In Section~\ref{sec:imaging_data} we give an
overview of the near-infrared (NIR) data acquisition and we describe
our 2-pass data reduction pipeline used to derive optimally (in terms
of signal-to-noise for faint sources) stacked images in
Section~\ref{sec:imaging_reduction}. We also present NIR galaxy number
counts and compare them with the literature.\\
In Section~\ref{sec:isel_catalogue} we present the deep multi-waveband
i-band selected catalogue and discuss its properties, whereas the data
reduction of the spectroscopic redshifts is described in
Section~\ref{sec:spec}.  In Section~\ref{sec:photoz} we present the
photometric redshift catalogue, discuss the accuracy of the latter and
show the redshift distribution of the galaxies.  In
Section~\ref{sec:uvlf} we derive the redshift evolution of the
restframe UV luminosity function and luminosity density at 1500~\AA\ 
from our i-selected catalogue before we summarise our findings in
Section~\ref{sec:summary_conclusion}.

We use AB magnitudes and adopt a $\Lambda$ cosmology throughout the paper
with \mbox{$\Omega_M=0.3$}, \mbox{$\Omega_\Lambda=0.7$}, and 
\mbox{$H_0=70 \, \mathrm{km} \, \mathrm{s}^{-1} \, \mathrm{Mpc}^{-1}$}.

\section{NIR observations}
\label{sec:imaging_data}

\subsection{Field layout}

Our observing strategy was designed to follow-up the publicly
available COSMOS observations with proprietary imaging in H band. The
whole area is covered by 25 pointings (15.4'x15.4' each) to a depth
adequate to the public NIR data. The field layout and nomenclature is
shown in Fig.\ref{fieldnomenclature}. The programme was carried out as
a joint effort between three large extragalactic survey projects
currently being pursued at the Centro Astron\'omico Hispano Alem\'an
(CAHA), on Calar Alto: ALHAMBRA \citep{moles:1}, MANOS-wide 
\citep{roeser:1,zatloukal:1}
and MUNICS-Deep (Goranova et al., in prep.). In
addition, as a part of the MUNICS-Deep project, in one of the
pointings (01c) we have already collected deep Js- and K'-band data.

\subsection{Data acquisition}

The observations in H, Js and K' bands were carried out using the NIR
wide-field imager
OMEGA2000\footnote{http://w3.caha.es/CAHA/Instruments/O2000/index2.html},
operating at the prime focus of the CAHA 3.5m telescope. OMEGA2000 is
equipped with a HAWAII-2 HgCdTe 2048x2048 array. The instrument pixel
scale is 0.45\arcsec per pixel, providing a field-of-view of
15.4\arcmin x 15.4\arcmin.

Here we present the H-band observations in 15 pointings (see
Fig.\ref{fieldnomenclature}), collected during 11 nights spanned over
3 observing campaigns: December 2004, February 2005, and March/April
2005. Each pointing was observed for a total of at least 3 ksec. The
individual exposure times were 3 sec, on-chip co-added to produce
single frames of 60 sec each. Except for 05b, at least 50 such frames
(depending on the weather conditions) were observed per pointing.

In addition, observations in Js and K' bands in 01c were collected
during 7 night in the observing campaigns November 2003 and February
2006. The total exposure time in Js-band data was 8.2 ksec with the
same individual exposure time scheme as for the H band. The K'-band
observations have a total of 7.7 ksec exposure time with individual
exposures of 2 sec, co-added internally into single frames of 30 sec
each.

All observations were done using dithering pattern consisting of
typically 20 positions shifted with respect to one another by 20\arcsec. The
consecutive dithering sequences were repeated with the same pattern
but with an offset in the origin.

The observing log presenting filters, pointings nomenclature,
coordinates, number of frames, and the total exposure time per
pointing are listed in Table~\ref{tab:obslog} .

\begin{table}
\caption{\label{tab:obslog} The 15 COSMOS patches. The table gives the
filter, the field name, the coordinates (right ascension $\alpha$ and
  declination $\delta$) for the equinox 2000, the number of frames, and
  the total exposure time.} 
\begin{tabular}{cccccc}
filter &  field  & $\alpha$ & $\delta$ &frames & exp. time\\
       &         & (2000)   &  (2000)  &             & [ksec]\\
\hline
  H & 01c  &  10:01:26.00 & +02:26:41.0   &  93 & 5.58 \\ 
    & 01d  &  09:59:31.24 & +02:26:41.0   &  81 & 4.86 \\ 
    & 02c  &  10:00:28.61 & +02:26:41.0   &  84 & 5.04 \\ 
    & 02d  &  09:58:33.85 & +02:26:41.0   & 100 & 6.00 \\ 
    & 03b  &  09:59:31.23 & +02:41:01.0   &  50 & 3.00 \\ 
    & 03c  &  10:01:25.99 & +02:12:21.1   &  50 & 3.00\\ 
    & 03d  &  09:59:31.25 & +02:12:21.1   &  50 & 3.00 \\ 
    & 04b  &  09:58:33.84 & +02:41:00.9   &  74 & 4.44 \\ 
    & 04c  &  10:00:28.60 & +02:12:21.0   &  49 & 2.94 \\ 
    & 04d  &  09:58:33.86 & +02:12:21.0   &  50 & 3.00 \\ 
    & 05a  &  10:01:25.98 & +01:58:01.1   &  49 & 2.94 \\ 
    & 05b  &  09:59:31.26 & +01:58:01.1   &  24 & 1.44 \\ 
    & 06a  &  10:00:28.62 & +01:58:01.0   &  70 & 4.20 \\ 
    & 07a  &  10:01:25.98 & +01:43:41.1   &  61 & 3.66 \\ 
    & 08a  &  10:00:28.63 & +01:43:41.1   &  48 & 2.88 \\ 
 Js & 01c  &  10:01:26.00 & +02:26:41.0   & 136 & 8.16 \\ 
 K' & 01c  &  10:01:26.00 & +02:26:41.0   & 258 & 7.74 \\ 
\end{tabular} 
\end{table}

\section{NIR imaging data reduction}
\label{sec:imaging_reduction}

In this section we present the NIR data reduction in the Js, H, and K'
bands. We develop and describe a 2-pass reduction pipeline optimised
for reducing and stacking images of different quality to get an
optimal signal-to-noise ratio for faint (sky dominated)
{pointlike} objects. We extract Js, H, and K' selected
catalogues and show the accuracy of our photometric calibration. Based
on these catalogues we compare the number counts in the three NIR
filters with number counts taken from the literature.

\subsection{Basic reduction}

{ For the basic reduction we use our own reduction pipeline
  (see Goranova~et~al., in prep. for a detailed description) based on
  the IRAF\footnote{"IRAF is distributed by the National Optical
    Astronomy Observatories, which are operated by the Association of
    Universities for Research in Astronomy, Inc., under cooperative
    agreement with the National Science Foundation."} external package
  XDIMSUM\footnote{Experimental Deep Infrared Mosaicing Software;
    XDIMSUM is a variant of the DIMSUM package developed by P.
    Eisenhardt, M. Dickinson, S.A. Stanford, and J. Ward. F.  Valdes
    (IRAF group); see also~\citet{stanford:1}.}.  We use a 2-pass
  reduction implementing object-masking for both flat-field and sky
  determination. The two passes are as follows: }

\begin{itemize}
\item First pass:
\begin{itemize}
\item[a)] Constructing the 1$^\textrm{st}$ science-flat,
\item[b)] Subtracting the sky,
\item[c)] Masking bad pixels and bad regions,
\item[d)] Rejecting cosmic rays,
\item[e)] Aligning the images, and
\item[f)] Stacking the images
\end{itemize}
\item Second pass:
\begin{itemize}
\item[g)] Constructing the object mask from the stacked image
\item[h)] Masking all objects in the raw frames using the object-mask,
  repeating pass 1 (step a and b) with masked objects and proceed as
  in the 1$^\textrm{st}$ pass for step c, d, e, and f until final
  stacking.
\end{itemize}
\end{itemize}

\begin{figure}
\centering
\includegraphics[angle=0, width=0.50\textwidth]{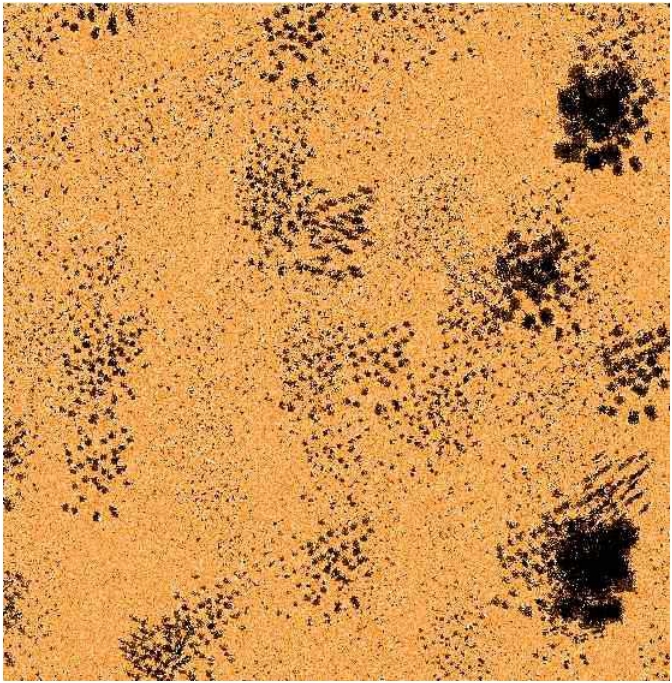}
\caption{Ratio of first and second pass flat-field: The first pass flat-field has been
  divided by the second pass flat-field. Dark regions (in the order of
  0.1\%) represent object residuals in the flat-field after the
  first pass. See text for details. \label{ratio_flat-field}}
\end{figure}

{The reasons for using this 2-pass reduction pipeline are the
  following: Due to severe problems with our dome and twilight sky
  flats (see Goranova~et~al. in prep.), we derive the flat-fields from
  the science frames themselves. In such a case, however, it is
  mandatory to to ensure that the final flat-field is free of object
  residuals.  Due to the very high noise level in the raw NIR data,
  this is possible only if all objects (both the bright and the faint
  ones) are masked out properly before deriving the flat-field, which
  we could not achieve with any other objects rejection algorithms
  ($\kappa-\sigma$ clipping, min-max rejection, etc.) we have applied.
  An example is presented in Fig.~\ref{ratio_flat-field}, where we
  show the ratio between the first pass (averaging after min-max
  rejection) and second pass (averaging after min-max rejection and
  object masking) flat-fields. The black regions clearly show the
  object residuals in the flat-field used for the first pass.
  Although the relative systematic error is in the order of only
  0.1\%, it makes a significant effect in the final stacked image. The
  reason for this is the very high sky level of the NIR images. In our
  case the sky level $N$ of a stacked image is at least in the order
  of 9 million photons. This implies, that the Poissonian error
  ($\sqrt N$) is about 3000 photons. If we now introduce a systematic
  error of 0.1\%, we end up with an error of 9000 photons. It is
  therefore clear that we introduce a systematic error of about $3
  \sigma$. For faint objects, which are detected on the level of a few
  $\sigma$ of the sky background, the systematic offset is in the same
  order as their total flux. In other words, if we introduce a
  systematic flat-field error of 0.1\%, the introduced systematic
  error is on the same level as the statistical Poissonian error for a
  sky level of 1 million photons.  }

\begin{figure}
\centering
\includegraphics[angle=0, height=0.29\textheight]{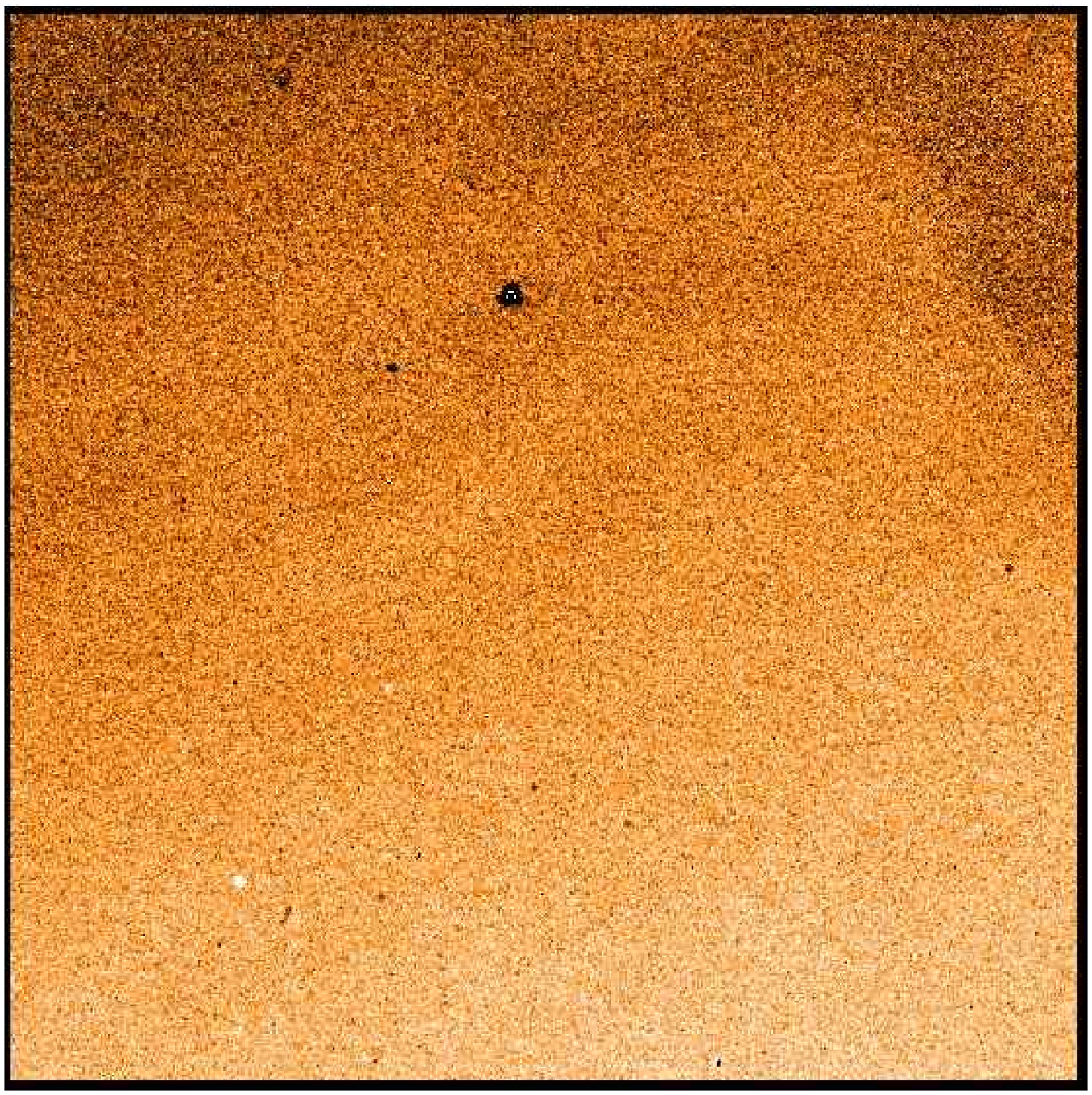}
\includegraphics[angle=0, height=0.29\textheight]{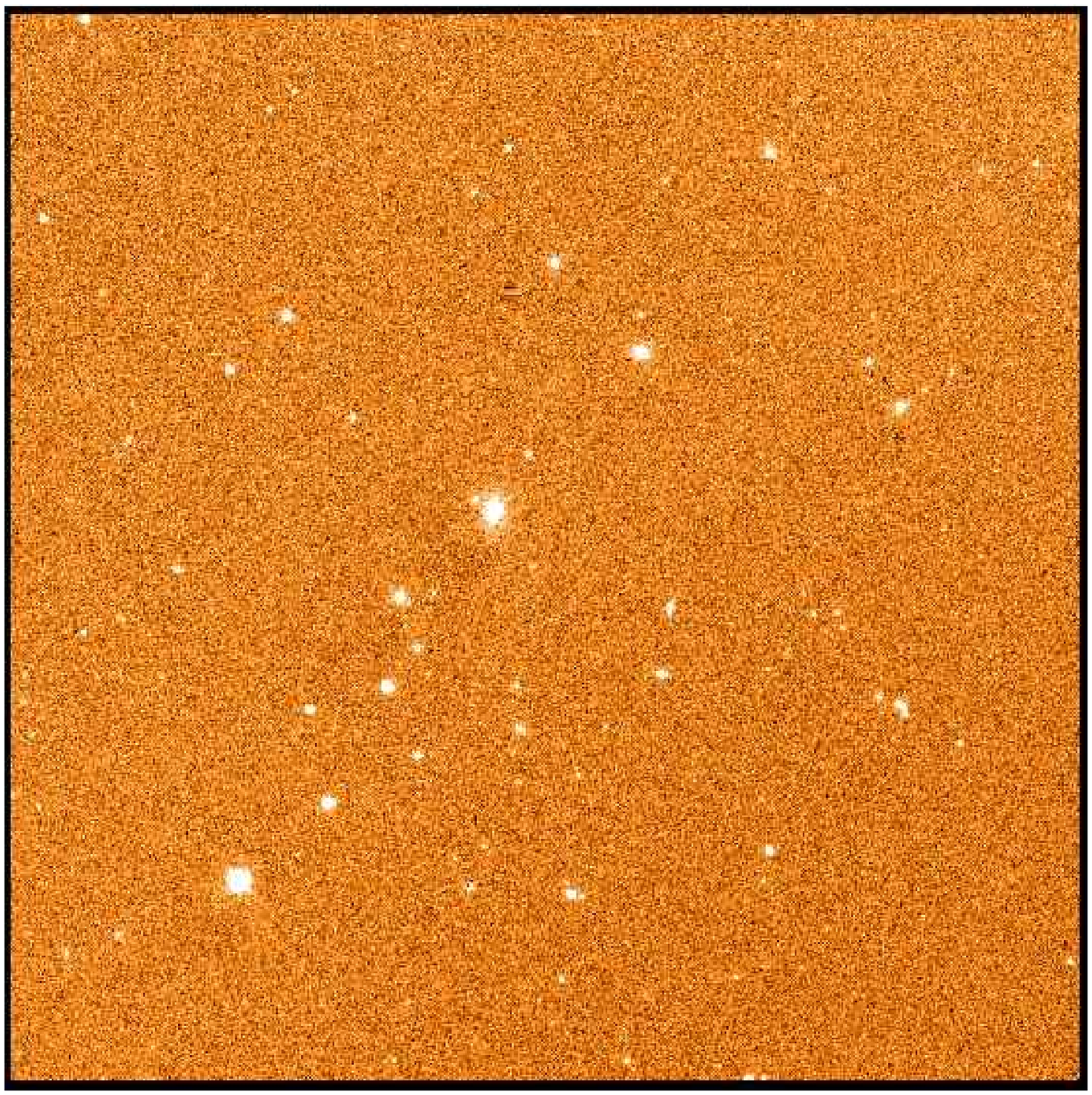}
\includegraphics[angle=0, height=0.29\textheight]{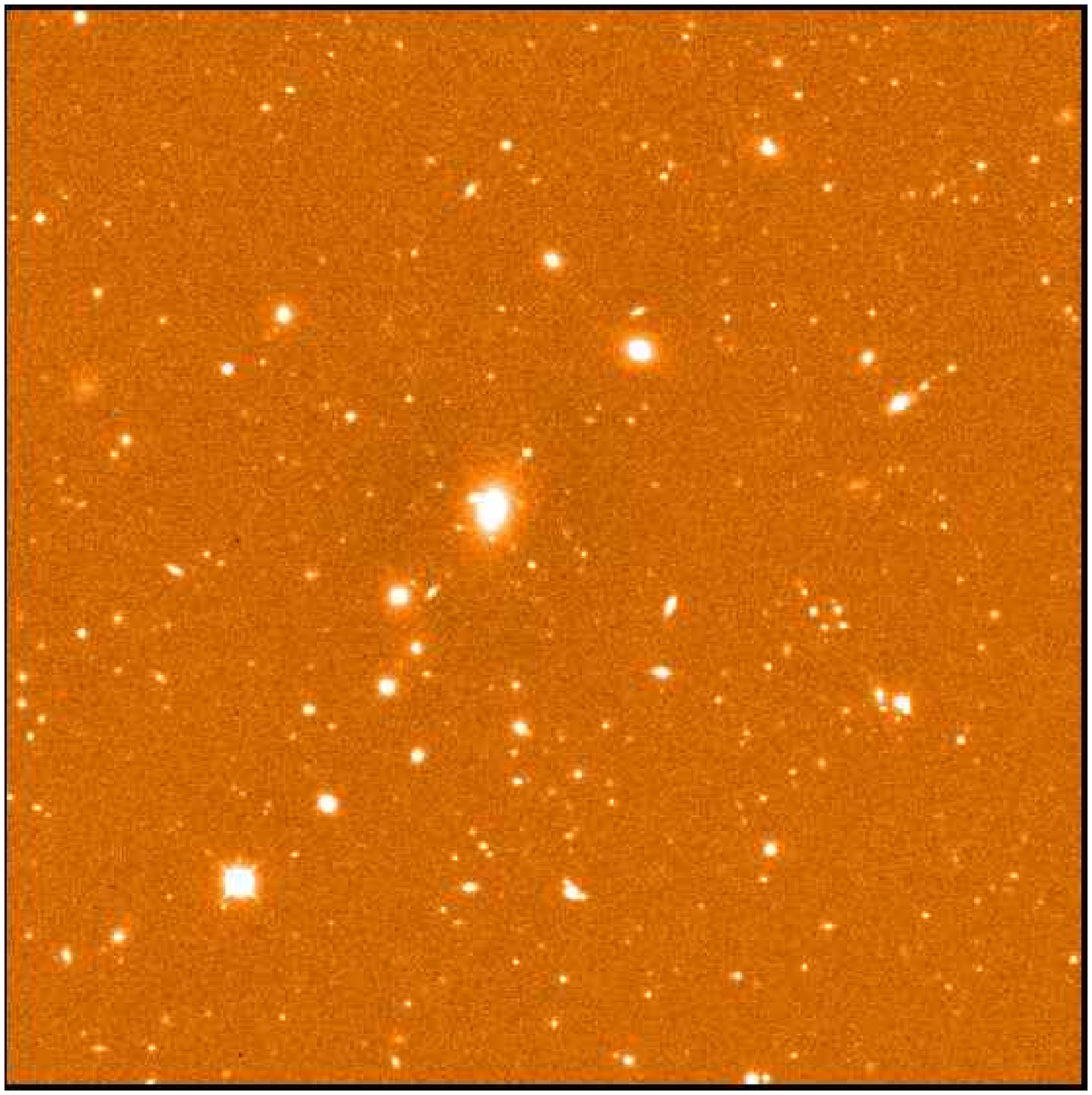}
\caption{
  Different steps of the reduction pipeline: A single raw H-band image
  (upper panel), the same image after pre-reduction (middle panel),
  and a stacked image after the two-pass data reduction and stacking
  with optimal S/N ratio (lower panel).
\label{pipeline_images}}
\end{figure}
\begin{figure*}\centering
  \includegraphics[angle=0,width=0.49\textwidth]{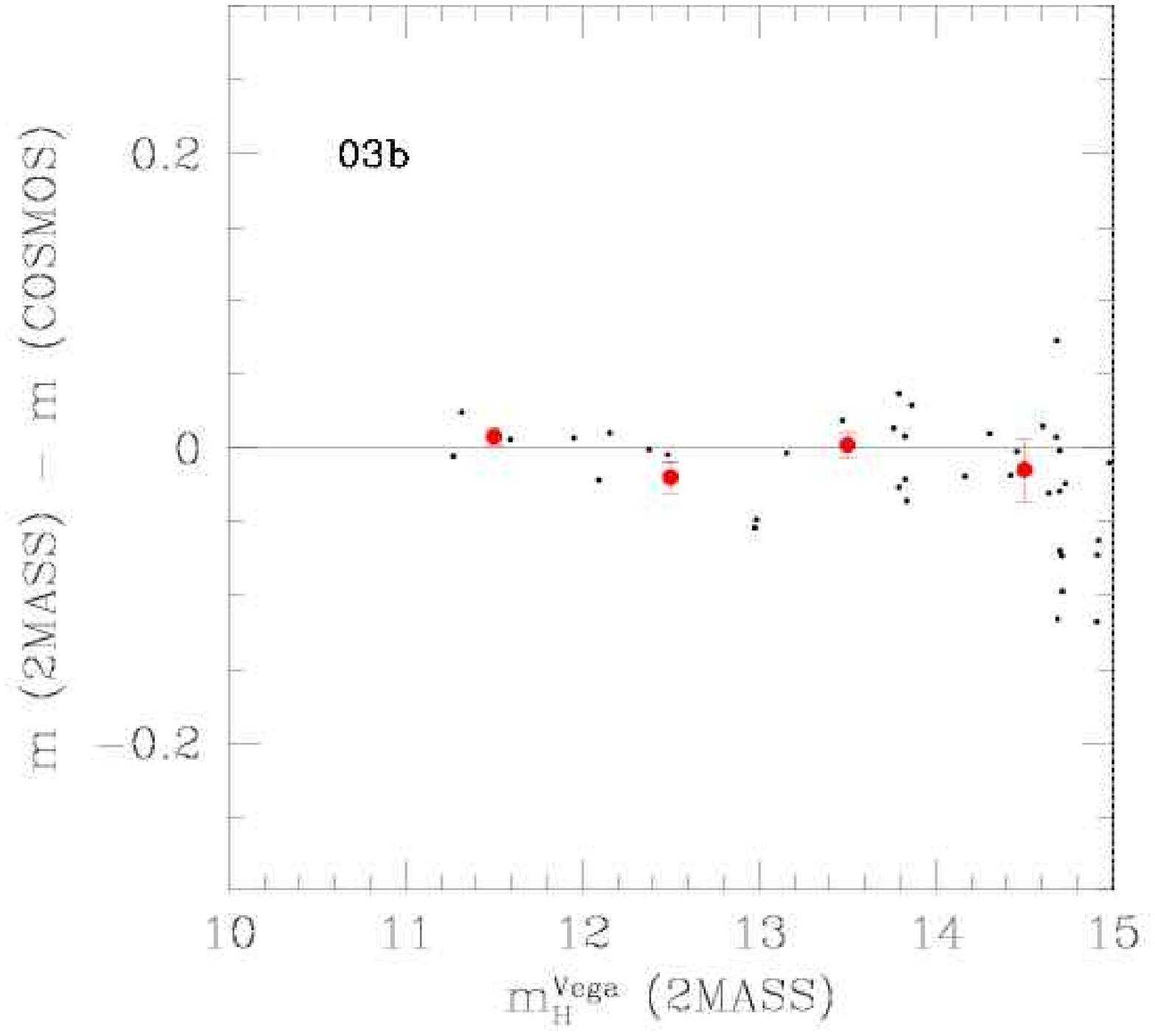}
  \includegraphics[angle=0,width=0.49\textwidth]{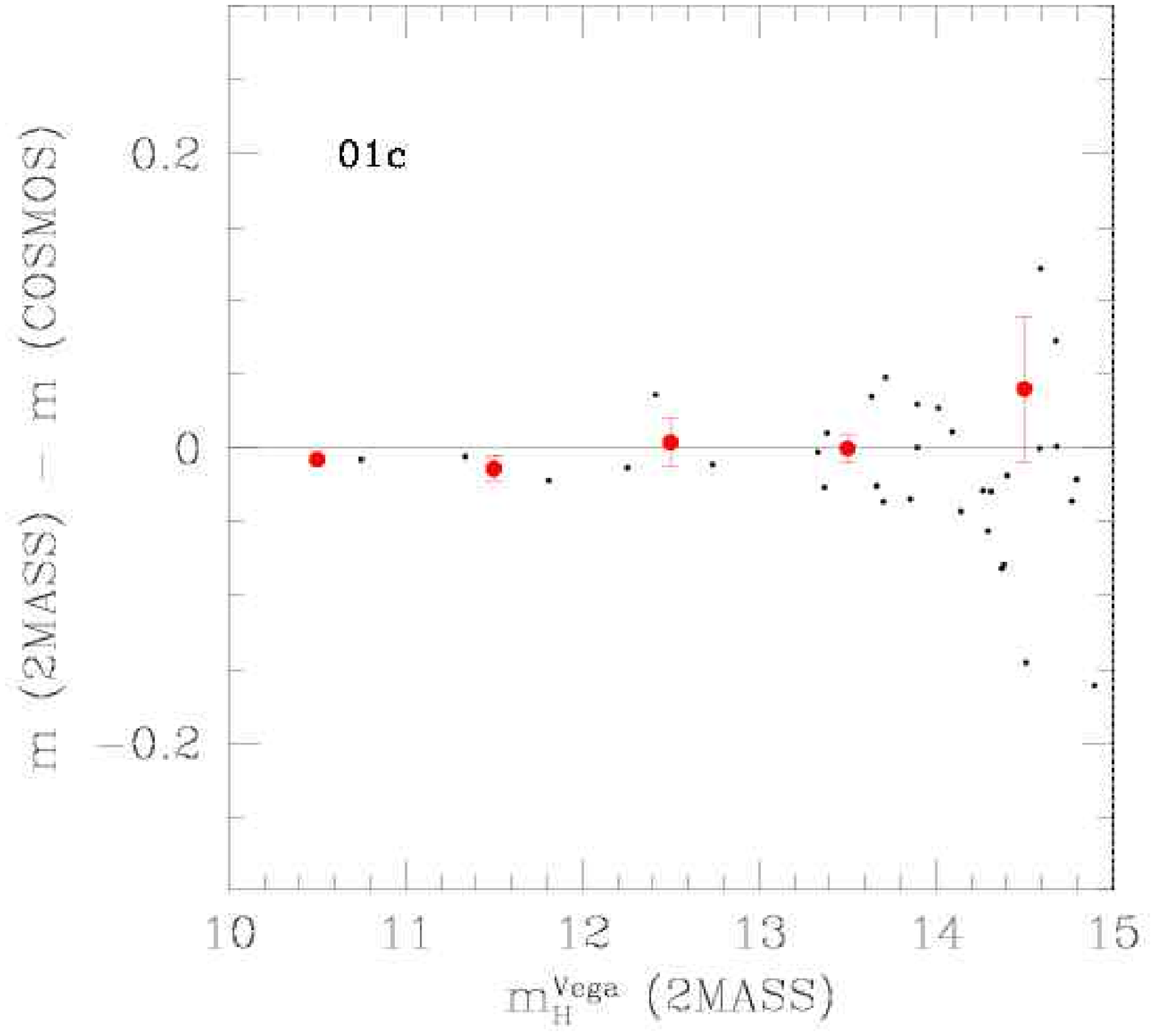}
\caption{H-band magnitude differences (black dots) between our COSMOS catalogue
  and 2MASS in 2 fields (03b and 01c). The red symbols show the
  average in the different magnitude bins. Please note that for these
  plots we used Vega magnitudes and not AB, in order to avoid adding
  any systematics by transforming the 2MASS magnitudes to AB system.
  The error in the zeropoint determination can be found in
  Table~\ref{tab:fieldnumbers}.
  \label{fig:calibration}}
\end{figure*}
\begin{figure*}
\centering
\includegraphics[angle=0,width=1.0\textwidth]{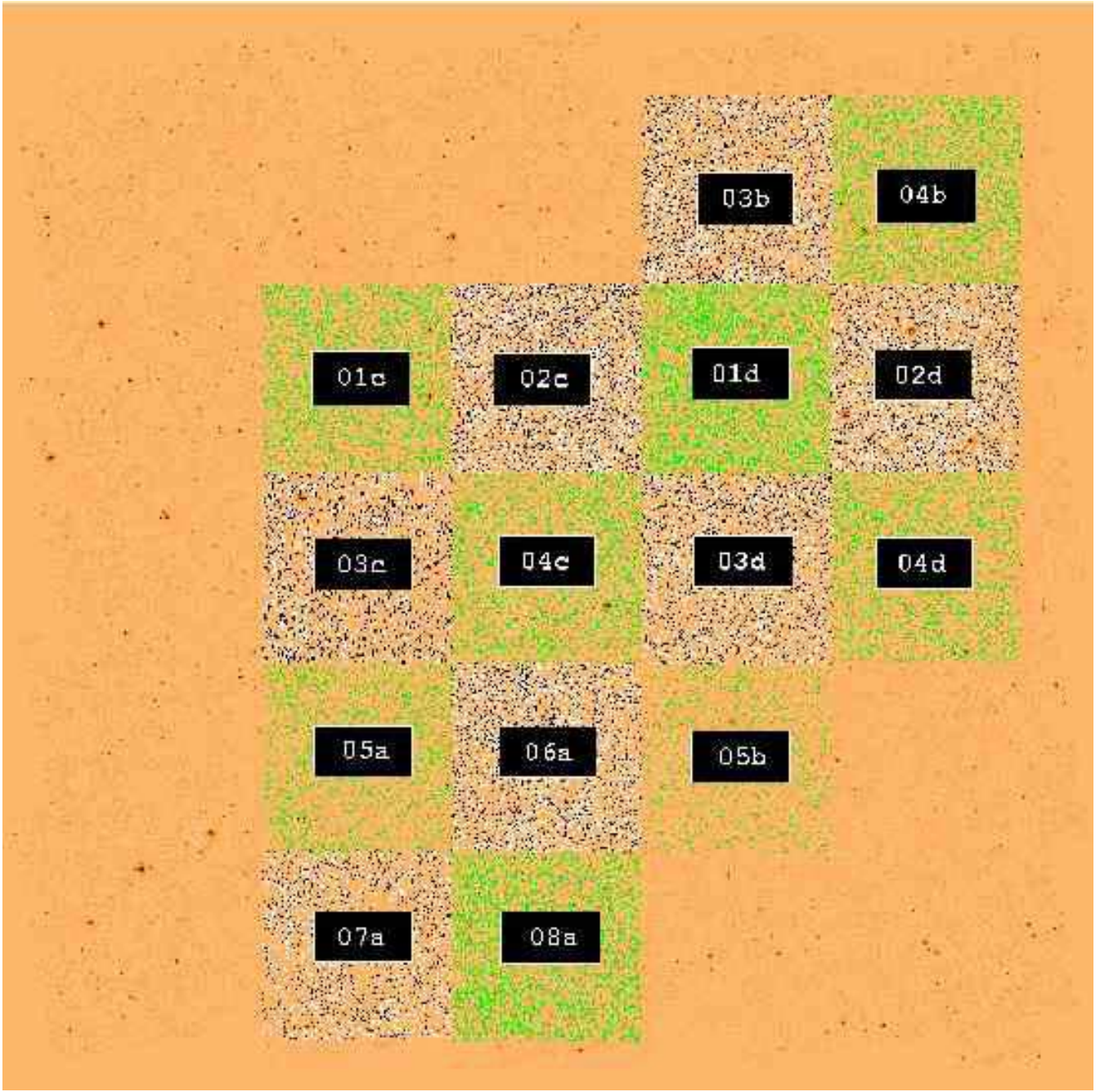}
\caption{Field nomenclature of the 15 NIR patches together with an excerpt of
  the H-band detected catalogues (green, black and white for clarity) on
  top of the HST COSMOS image. Each catalogue covers an
  area of $205.44\sq\arcmin$. \label{fieldnomenclature}}
\end{figure*}
\begin{table}
\caption{\label{tab:fieldnumbers} Field characteristics. The first
  column gives the filter band, the second the field nomenclature as
  used in Fig.~\ref{fieldnomenclature}. In the third column, the number of detected object
  is reported (see text for details). The 50\% completeness limit
  follows in the fourth column, the FWHM seeing value in the fifth, and
  the error of the photometric zero point in the last column.}

\begin{tabular}{c|c|c|c|c|c}\\
filter &  field  & objects & 50\% CL.&  seeing   & zeropoint error\\
       &         &    & [mag]      & [\arcsec] &  [mag]   \\
\hline
  H & 01c   & 2744 & 21.76     &  1.13  & 0.006  \\ 
    & 01d   & 4013 & 22.38     &  0.82  & 0.005  \\ 
    & 02c   & 3483 & 22.24     &  0.92  & 0.007  \\ 
    & 02d   & 3364 & 22.40     &  0.98  & 0.008  \\ 
    & 03b   & 2996 & 21.82     &  1.14  & 0.006  \\ 
    & 03c   & 2020 & 21.27     &  1.41  & 0.007  \\ 
    & 03d   & 2765 & 21.87     &  1.24  & 0.007  \\ 
    & 04b   & 2846 & 22.25     &  1.00  & 0.006  \\ 
    & 04c   & 2567 & 21.68     &  1.09  & 0.006  \\ 
    & 04d   & 2348 & 21.52     &  1.39  & 0.008  \\ 
    & 05a   & 1815 & 21.16     &  1.07  & 0.008  \\ 
    & 05b   & 1102 & 20.60     &  1.56  & 0.006  \\ 
    & 06a   & 2981 & 22.16     &  0.83  & 0.008  \\ 
    & 07a   & 1940 & 21.52     &  0.80  & 0.007  \\ 
    & 08a   & 2618 & 21.83     &  0.94  & 0.019  \\ 
  Js & 01c  & 4005 & 22.67     &  0.91  & 0.009  \\ 
  K' & 01c  & 3174 & 21.76     &  1.05  & 0.009  \\ 
\end{tabular}
\end{table}

\begin{figure}
  \centering \includegraphics[angle=0,width=0.50\textwidth]{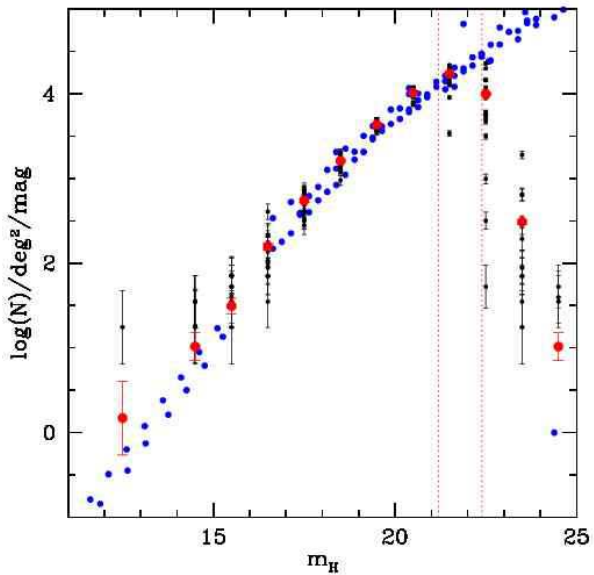}
\caption{
  Galaxy number counts in the H-band (not corrected for
  incompleteness) as compared to the literature (blue dots).  The
  black dots represent the number counts of the single patches,
  whereas the red dots show the average number counts of all fields.
  The vertical dotted lines indicates the 50\% completeness limits of
  the shallowest (05a) and the deepest (02d) patches. The literature
  number counts are taken from
  \citet{yan:2,teplitz:3,thompson:1,martini:1,chen:2,moy:1,firth:1,metcalfe:2}.
  The error bars show the $1\sigma$ Poissonian errors.
\label{nc_all_H}
}
\end{figure}
\begin{figure*}\centering
  \includegraphics[angle=0,width=0.49\textwidth]{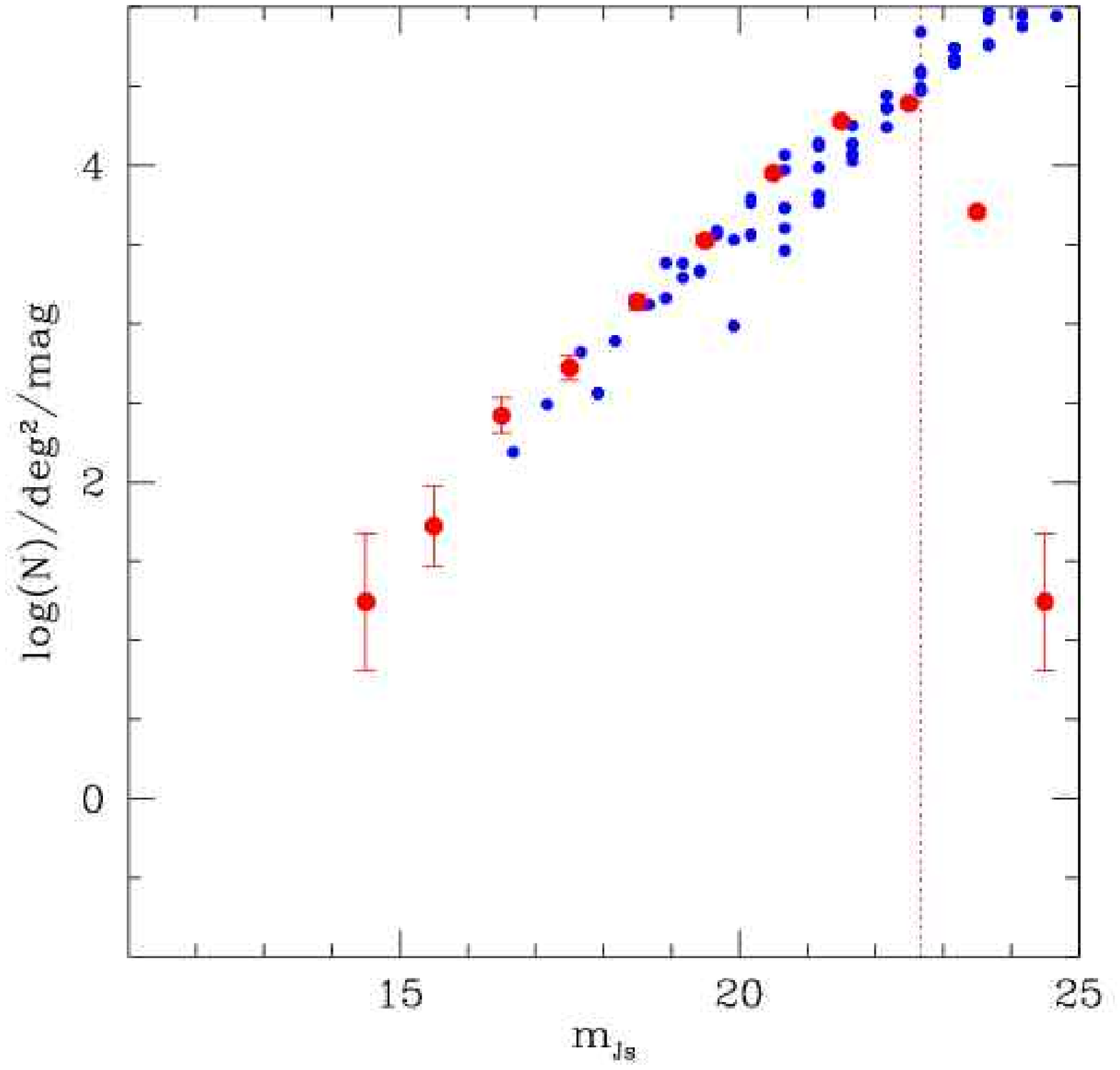}
  \includegraphics[angle=0,width=0.49\textwidth]{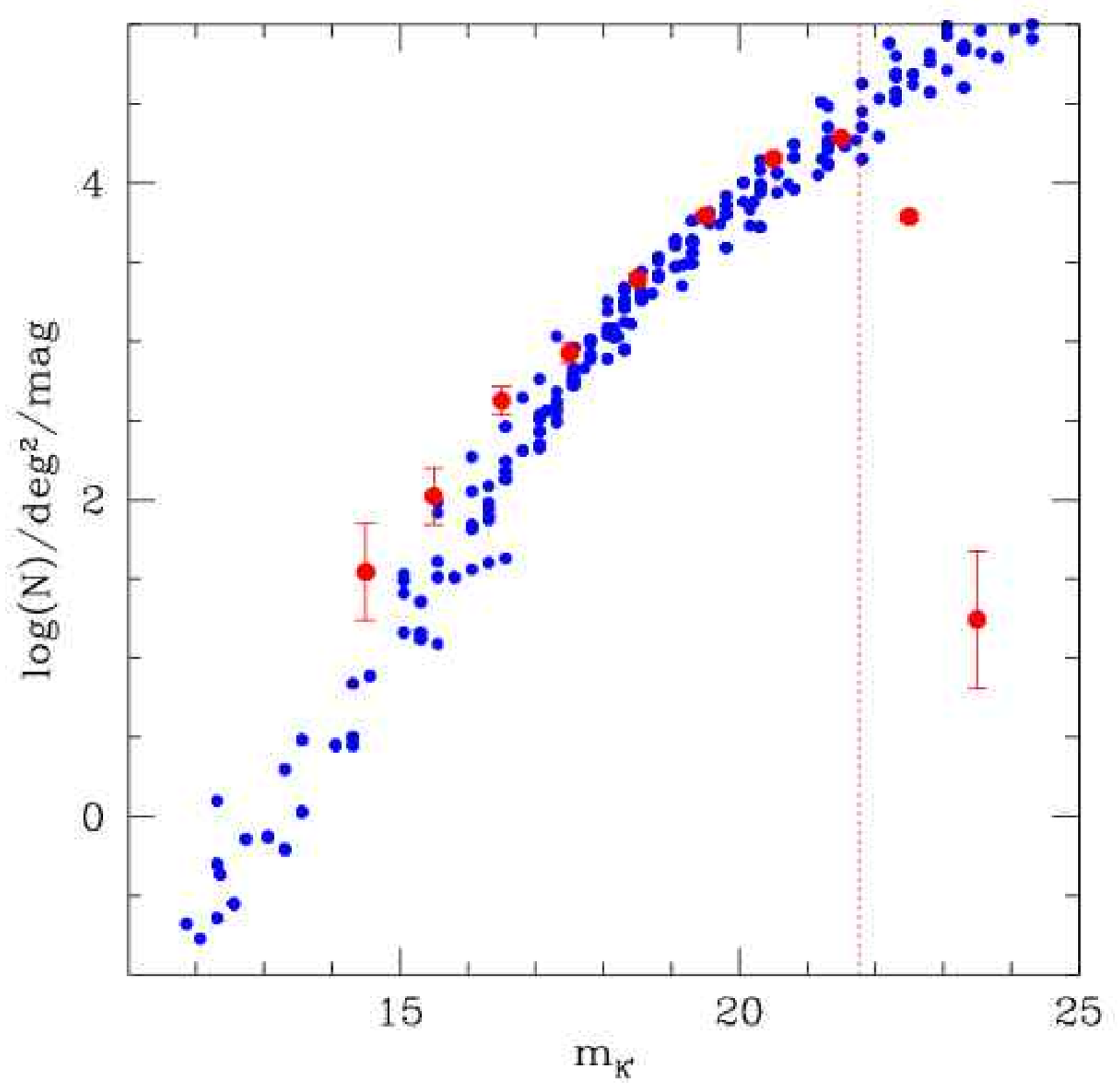}
\caption{ Left panel: Js-band galaxy number counts 
  of patch 01c (red dots, not corrected for incompleteness) as
  compared to the literature (blue dots).  The vertical dotted line
  indicates the 50\% completeness limit. The
  literature number counts are taken from \citet{Saracco99,
saracco2001, maihara2001s, Teplitz99}. Right panel: K'-band galaxy
  number counts of patch 01c (red dots, not corrected for
  incompleteness) as compared to the literature (blue dots).  The
  vertical dotted line indicates the 50\% completeness limit.  The
literature number counts are taken from \citet{gardner1993,
soifer1994, djorgovski1995, glazebrook1995, mcleod1995, gardner1996,
huang1997, moustakas1997, saracco1997, bershady1998, szokoly1998,
mccracken2000, vaisanen2000, munics1, huang2001, kuemmel2001,
maihara2001s, martini2001, saracco2001, cristobal2003, minowa2005}.
 The error bars show the $1\sigma$ Poissonian errors.
  \label{nc_01c_JK}}
\end{figure*}

{ Another problem of NIR data reduction is the short-scale time
  variability of the sky. We subtract the sky using the so-called
  'running' sky technique, i.e we determine the sky for each frame
  individually. For this we take a series of usually 5 frames grouped
  in time before and after (but excluding) the frame we are
  subtracting the sky from and we average those using min-max
  rejection and object masking. This works very well as the
  sky-illumination does not change much within this short time.\\
  The variability of the sky on the other hand may change the overall
  shape of the flat-field (derived before sky subtraction). In order
  to avoid correcting this additive illumination by the multiplicative
  flat-field, we use the following approach. First we derive
  flat-fields for each night by averaging all frames from that night
  using min-max rejection and object masking. Then we average the
  normalised flat-fields from all nights in order to derive a master
  flat-field. This master flat-field should be the best approximation
  to the real shape of the overall sky-flat. Then we fit a
  4$^{th}$-order Chebyshev polynomial to the master flat-field in
  order to get the shape of the master flat-field and eliminate the
  pixel-to-pixel variations. Finally, we bend every single flat-field
  to the shape of the Chebyshev flat-field. The corrections applied to
  the single flat-fields are in the order of 1--2\% (peak to peak).  }

Since our final goal is the combination of our H-band data with public
data to construct a multi-wavelength catalogue, we want to
astrometrically align our H-band images to the publicly available
images in the COSMOS field \citep{scoville:1}. Moreover, as
we intend to use the same procedure as in the FDF in constructing the
catalogue \citep{fdf_data}, i.e. using SExtractor in the double image
mode, the data sets in the different passbands must also be aligned by
pixel. On the other hand, we do not want to shift the NIR images
twice. This would smooth the images and destroy the signal of very
faint objects.  As it is also very challenging to shift single NIR
images on top of deep optical images (where many bright starlike
objects are e.g.  saturated), we use the
following approach:\\
First we compute the astrometric transformation between our stacked
NIR image and the reference optical frame (Subaru z-band). This allows
us to rely on hundreds of objects when computing the astrometric
solution instead of only a few tenths of objects visible in a single NIR
image.  Knowing the relative shifts of the single images to the
stacked XDIMSUM image and the full astrometric solution of the latter
(with respect to the Subaru z-band), we are able to calculate the
astrometric solution also for each single, \textit{unshifted}
pre-reduced image. Restarting from these images, we do the alignment,
the distortion correction, the regridding and the stacking of the
individual images (with optimal signal-to-noise ratio, see below) in
\textit{one} step.

\subsection{Image stacking with optimal S/N for point sources}

As the single exposures were not taken under the same observing
conditions, the sky levels, the seeing, and the zeropoints can
substantially differ from frame to frame. On the other hand this gives
one the possibility to stack the single images with weighting factors
in order to achieve a final combined frame with optimal
signal-to-noise ratio (S/N) {for point sources}.\\ We
calculated a weight $\alpha$ to be applied to an individual image
following the general \textit{Ansatz} for two images (denoted by index
1 and index 2):
\begin{eqnarray}
\label{eqn:datareduction:ansatz}
S_{tot}& = &S_1+\alpha_2 S_2 \nonumber \\
N_{tot}& = &\sqrt{N_1^2+(\alpha_2 N_2)^2}
\end{eqnarray}
where $S_{tot}$ and $N_{tot}$ is the signal and the noise of the
combined image. This transforms to:

\begin{equation}\small
\label{eqn:datareduction:observable}
\frac{S_{tot}}{N_{tot}}=\frac{f_1+\alpha_2 f_2}
{\sqrt{(f_1+h_1\sigma_1^2\pi)+\alpha_2^2(f_2+h_2\sigma_2^2\pi)}} 
\end{equation}
where 
$f_1$  and $f_2$ are the fluxes of an object without sky (signal),
$h_1$  and $h_2$ are the sky values and
$\sigma_1$ and $\sigma_2$ correspond to the seeing in the two frames.
$\alpha_2$  is the  weighting factor to be applied to frame~2.
It is than straight forward to compute the value of $\alpha_2$ for
which $S_{tot}/N_{tot}$ is maximised:
\begin{equation}
\label{eqn:datareduction:alpha}
\frac{\partial\frac{S_{tot}}{N_{tot}}}{\partial \alpha_2 }\stackrel{!}{=}0 \Rightarrow
\alpha_2=\frac{f_2(f_1+h_1\sigma_1^2\pi)}{f_1(f_2+h_2\sigma_2^2\pi)}
\end{equation}
For bright objects which are not dominated by the sky noise ($f\gg h$)
Equation~(\ref{eqn:datareduction:alpha}) approaches the limit
\begin{equation}
\label{eqn:datareduction:alpha_bright}
\alpha_2=1,
\end{equation}
whereas for faint, sky dominated objects ($f\ll h$)
Equation~(\ref{eqn:datareduction:alpha}) transforms into
\begin{equation}
\label{eqn:datareduction:alpha_faint}
\alpha_2=\frac{f_2\cdot(h_1\sigma_1^2)}{f_1\cdot(h_2\sigma_2^2)}
\end{equation}
As the overwhelming majority of the objects are very faint point
sources and therefore dominated by the sky noise,
Equation~(\ref{eqn:datareduction:alpha_faint}) is used to derive the
weighting-parameter $\alpha$.

To reduce errors when determining the weighting-parameter $\alpha$, the
fluxes $f_{1/2}$ are derived from a bright not saturated star, the sky
levels $h_{1/2}$ correspond to the mode\footnotemark\footnotetext{most
  frequent value in the pixel histogram of the image} of the image and
the seeings $\sigma_{1/2}$ have been calculated from the median seeing
of a few stars.

The weighting factor for the $1^{\rm st}$ frame (randomly chosen) was
set to unity. The factors for all other \mbox{(N $-$ 1)} images were
then derived following Equation~(\ref{eqn:datareduction:alpha_faint})
relative to this image. The final stacked frame $I_{sum}$ can then be
calculated according to:
\begin{equation}\large
I_{sum}=\sum_{i=1}^N \alpha_i \ B_i \ \tilde{I_i} \cdot 
\frac{\sum_{i=1}^N \alpha_i \ f_i}{\sum_{i=1}^N \alpha_i \ f_i \ B_i} 
\label{eqn:datareduction:stack}
\end{equation}
where the index $i$ denotes a single image.  $\alpha_i$ is the
weighting factor according to
Equation~(\ref{eqn:datareduction:alpha_faint}); $\tilde{I_i}$ is the
sky-subtracted single image; $B_i$ is the bad pixel
mask\footnotemark\footnotetext{zero for a bad pixel, unity otherwise}
and $f_{i}$ is the flux derived from the bright star (used to derive
$\alpha_i$; see also Equation~\ref{eqn:datareduction:observable}).

Finally, we did the image alignment, the distortion correction, the
regridding and the stacking of the individual images with optimal S/N
in one step by using standard IRAF routines as well as SWarp
\citep{bertin:1}. The regridding has been done to the native pixel
scale of the Subaru telescope (0.2\arcsec \ per pixel). Please note
that we do not interpolate any bad pixel or bad region in the single
images, but set them to zero during the stacking procedure. Missing
flux in bad regions is taken into account by using
Equation~(\ref{eqn:datareduction:stack}).\\
Since a different number of dithered frames contributed to each pixel
in the co-added images (producing a position-dependent noise pattern)
a combined weight map for each frame was constructed.  The latter was
used during source detection and photometry procedure to properly
account for the position-dependent noise level.

In Fig.~\ref{pipeline_images} we illustrate different steps of our
reduction pipeline: a single raw H-band image, the same image after
pre-reduction, and a stacked image after the two-pass data reduction
and stacking with optimal S/N ratio. There are practically no haloes
around very bright objects, e.g.\ due to object residuals in the
flat-field or sky subtraction problems~/~residuals.

\subsection{Photometric calibration}

The absolute photometric calibration is based on the 2MASS
\citep{2MASS} catalogue. First we cross-correlate the objects in all
our reduced patches with sources in the 2MASS catalogue. We exclude
all objects with possible problems in either one of the catalogues
relying on the quality flags of 2MASS and SExtractor quality flags
(only objects with a flag of $\le 3$ are considered). An error
weighted fit between these objects (20 to 45, depending on the patch)
then determines the zeropoint as well as its error. In
Fig.~\ref{fig:calibration} we show typical magnitude differences
between our COSMOS catalogue and 2MASS.  Please note that for these
plots we used Vega magnitudes and not the AB magnitude systems (we did
not want to add any systematic by changing the 2MASS magnitude
system). The accuracy of the zeropoint in the different fields is in
the order of 0.01 magnitudes (derived from the error weighted fit
between the COSMOS and 2MASS objects; see above) and can be found in
Table~\ref{tab:fieldnumbers}.

\subsection{NIR catalogues}

Based on the stacked images we derived 15 H-band selected catalogues
(see Fig.~\ref{fieldnomenclature}) as well as 1 J-band and 1 K-band
selected catalogue (in field 01c, only). For this purpose we run
SExtractor \citep{bertin} with the detection threshold $t=2$ (minimum
signal-to-noise ratio of a pixel to be regarded as a detection) and
$n=3$ (number of contiguous pixels exceeding this threshold).
Depending on the depth\footnote{as a result of the slightly varying
  seeing and total exposure times, see Table~\ref{tab:fieldnumbers}}
of the different patches, we detect between 2000 and 4000 objects
(excluding patch 05b were we have only half of the minimum exposure
time we wanted to achieve). The false detection rate are in the order
of one percent or less (detected on the inverted image). Only around
extremely bright and saturated objects (three in patch 02d and one in
patch 03c) the false detection rate increases. As the regions around
these four objects are not taken into account, we get an overall
false detection rate in the order of 1 percent.\\
Because the depth of the patches decreases towards the borders, we
limited our analysis to the inner field. The signal-to-noise ratio in
this `deep' region is at least 90\% of the best S/N in every patch.
This prevents a possible bias of the photometric redshifts due to a
not completely homogeneous data set.  The single patches, each
covering an area of $205.44\sq\arcmin$ together with the field
nomenclature are shown in Fig.~\ref{fieldnomenclature}. In total we
detected about 40~000 H-selected objects over an area of about
$0.85\sq^{o}$ as well as about 4000 (3000) J (K) selected galaxies
over an area of $205.44\sq\arcmin$.

\subsection{NIR number counts}

The galaxy number counts can be used to check the calibration of the
data set, to detect possible galaxy over or under-densities of a field
as well as to determine the approximate depth of the data. We did not
put much effort in star-galaxy separation at the faint end (as we did
for the i-selected catalogue, see Sect.~\ref{sec:isel_catalogue}),
where the galaxies dominate the counts.  At the bright end, where
SExtractor is able to disentangle a stellar and a galaxy profile, we
used the star-galaxy classifier to eliminate obvious stellar objects.
We present in Fig.~\ref{nc_all_H} the H-band number counts of all
patches and in Fig.~\ref{nc_01c_JK} the Js and K' band number counts
of patch 01c.  Although the single patches show a scatter in the
H-band number counts at the very bright end, there is a good to very
good agreement between the literature number counts and our mean
number counts (red dots) up to the 50 \% completeness limit for point
sources \citep{snigula:1}.  Comparing the Js and K' band number counts
with data taken from the literature (Fig.~\ref{nc_01c_JK}) also shows
a relatively good agreement, although patch 01c seems to be overdense
with respect to most of the literature values. This is also true for
the H-band number counts in 01c if compared to the other patches or to
the literature, indicating an overdensity in this specific patch.
Nevertheless, our results are compatible within the error bars at a
$1\sigma$ level.\\ Please note that the 50 \% completeness levels for
point sources (vertical red lines in Fig.~\ref{nc_all_H} and
Fig.~\ref{nc_01c_JK}) coincide very nicely with the faint-end region
where the number counts start do drop. The values of the NIR galaxy
number counts can be found in Table~\ref{tab:nc}.

\begin{table*}

\caption{Galaxy number counts not corrected for incompleteness from COSMOS in the
$i$ (12 patches), $Js$ (1 patch), $H$ (15 patches), and $K'$ (1 patch) bands. $\log N$ and $\sigma_{\log N}$ are 
given, where $N$ is in units of $\mathrm{mag}^{-1} \mathrm{deg}^{-2}$.}
\label{tab:nc}
\begin{center}
\begin{tabular}{ccccccccc}
\hline
& $i$ & & $Js$ & & $H$ & & $K'$  \\
$m$ & $\log N$ & $\sigma_{\log N}$ & $\log N$ & $\sigma_{\log N}$ & $\log N$ &
$\sigma_{\log N}$ &$\log N$ & $\sigma_{\log N}$  \\\hline

 14.5 &          &        &    1.244  &   0.434   &  1.015  &    0.164    &        1.545  &   0.307  \\      
 15.5 &          &        &    1.721  &   0.250   &  1.492  &    0.094    &        2.022  &   0.177  \\      
 16.5 &    0.465 &  0.307 &    2.420  &   0.112   &  2.195  &    0.042    &        2.624  &   0.088  \\      
 17.5 &    1.465 &  0.097 &    2.721  &   0.079   &  2.738  &    0.022    &        2.925  &   0.062  \\      
 18.5 &    2.304 &  0.036 &    3.136  &   0.049   &  3.206  &    0.013    &        3.393  &   0.036  \\      
 19.5 &    3.127 &  0.014 &    3.525  &   0.031   &  3.631  &    0.008    &        3.790  &   0.023  \\      
 20.5 &    3.598 &  0.008 &    3.949  &   0.019   &  4.010  &    0.005    &        4.152  &   0.015  \\      
 21.5 &    3.996 &  0.005 &    4.277  &   0.013   &  4.236  &    0.004    &        4.283  &   0.013  \\      
 22.5 &    4.350 &  0.003 &    4.390  &   0.011   &  3.995  &    0.005    &        3.784  &   0.023  \\      
 23.5 &    4.677 &  0.002 &    3.705  &   0.025   &  2.486  &    0.030    &        1.244  &   0.434  \\      
 24.5 &    4.970 &  0.001 &    1.244  &   0.434   &  1.015  &    0.164    &               &          \\      
 25.5 &    5.103 &  0.001 &           &           &  0.170  &    0.434    &               &          \\
 26.5 &    4.983 &  0.001 &           &           &         &             &               &          \\           
 27.5 &    4.338 &  0.003 &           &           &         &             &               &          \\   
 28.5 &    2.702 &  0.023 &           &           &         &             &               &          \\   
 29.5 &    1.278 &  0.120 &           &           &         &             &               &          \\\hline   

\end{tabular}
\end{center}

\end{table*}

\section{The i-selected catalogues}
\label{sec:isel_catalogue}

Based on the publicly available optical and NIR data of the COSMOS
field\footnote{The data were taken from:\\
  http://irsa.ipac.caltech.edu/data/COSMOS/} we build a Subaru i-band
detected galaxy catalogue in 12 of our 15 patches. Because of the
relatively bad H-band seeing ($>1.3$\arcsec), we exclude the patches
03c, 04d as well as 05b.
\begin{figure*}\centering
  \includegraphics[angle=0,width=0.49\textwidth]{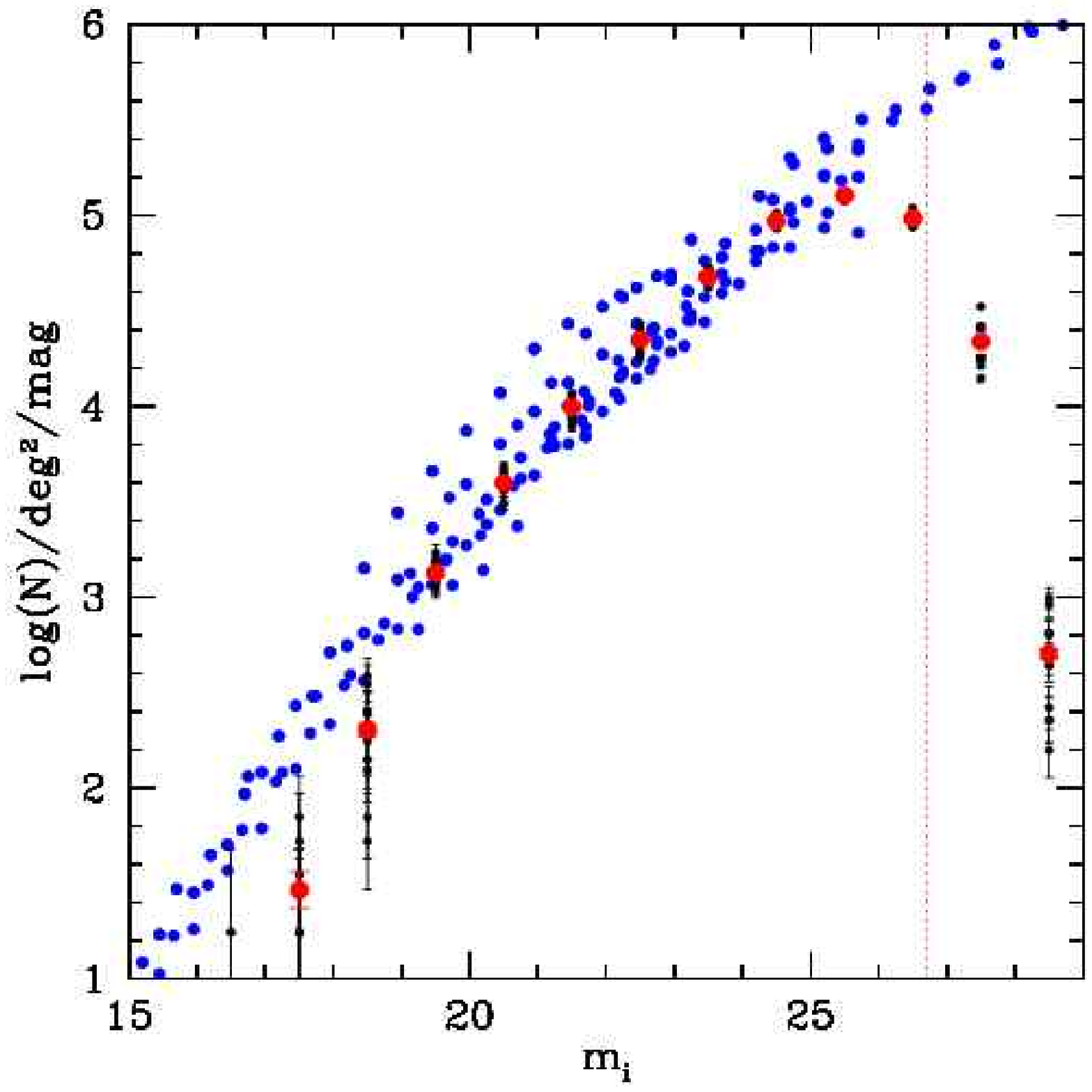}
  \includegraphics[angle=0,width=0.49\textwidth]{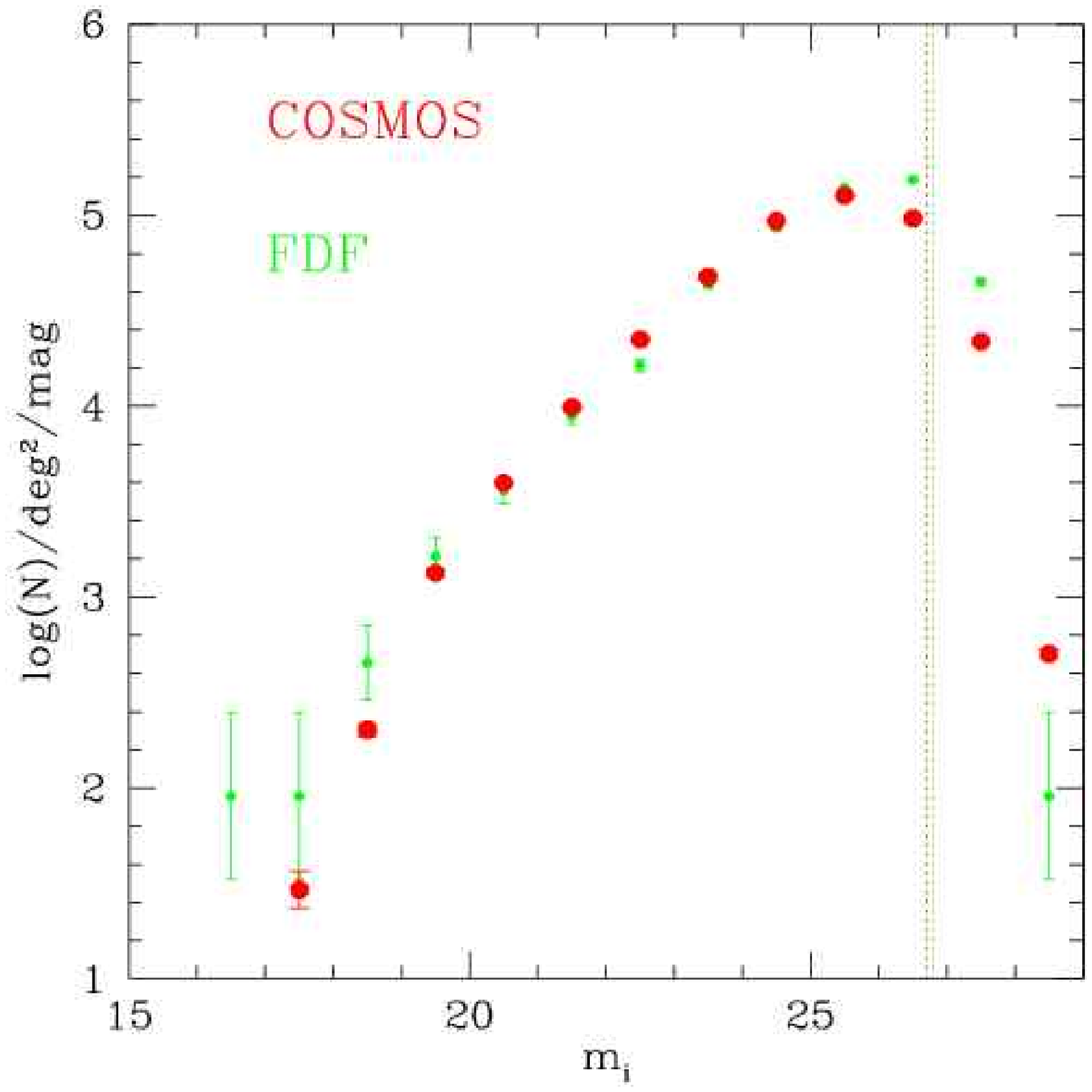}
\caption{  Left panel: Galaxy number counts in the i-band (not corrected for
  incompleteness) as compared to the literature. The black dots
  represent the number counts of the single COSMOS patches, whereas
  the red dots show the average number counts of all fields. The
  vertical dotted line indicates the 50\% completeness limit. The
  literature number counts are taken from \citet{HM84, Tyson88, LCG91,
CRGINOW95, DWOKGR95, HDFN, HCL98, PL98, ADCZFG99, MSCMF2001,
EISCDFS2001, Yasuda2001s, Capak2004}.
Right panel: Galaxy number counts in the i-band (not
  corrected for incompleteness) as compared to the FDF. The red dots
  show the average number counts of all COSMOS fields, whereas the
  green dots show the number counts as derived from the deep part of
  the FDF.  The vertical dotted lines indicate the 50\% completeness
  limits of COSMOS and FDF. The error bars show the
  $1\sigma$ Poissonian errors.\label{nc_all_i}}
\end{figure*}
\begin{table}
\caption{\label{tab:public_data} 
COSMOS (FDF) field characteristics for a SExtractor detection threshold of $t=2.5$ ($t=1.7$) and
  $n=3$ ($n=3$)
  contiguous pixels.}
\centering
\begin{tabular}{c|c|c|c}\\
Filter & 50 \% CL.   & 50 \% CL. &  seeing   \\
       & COSMOS      & FDF       & [\arcsec] \\
\hline
  u$_{CFHT}   $ &     25.6  & 26.5 (U)   &  0.90   \\
  B$_{Subaru} $ &     27.7  & 27.6 (B)   &  0.95   \\
  V$_{Subaru} $ &     26.5  & 26.9 (g)   &  1.30   \\
  r$_{Subaru} $ &     26.8  & 26.9 (R)   &  1.05   \\
  i$_{Subaru} $ &     26.7  & 26.8 (I)   &  0.95   \\
  z$_{Subaru} $ &     25.1  & 25.8 (z)   &  1.15   \\
  Ks$_{KPNO}  $ &     21.2  & 22.6 (Ks)  &  1.28   \\
\end{tabular}
\end{table}
\begin{figure*}
\centering
  \includegraphics[angle=0,width=0.49\textwidth]{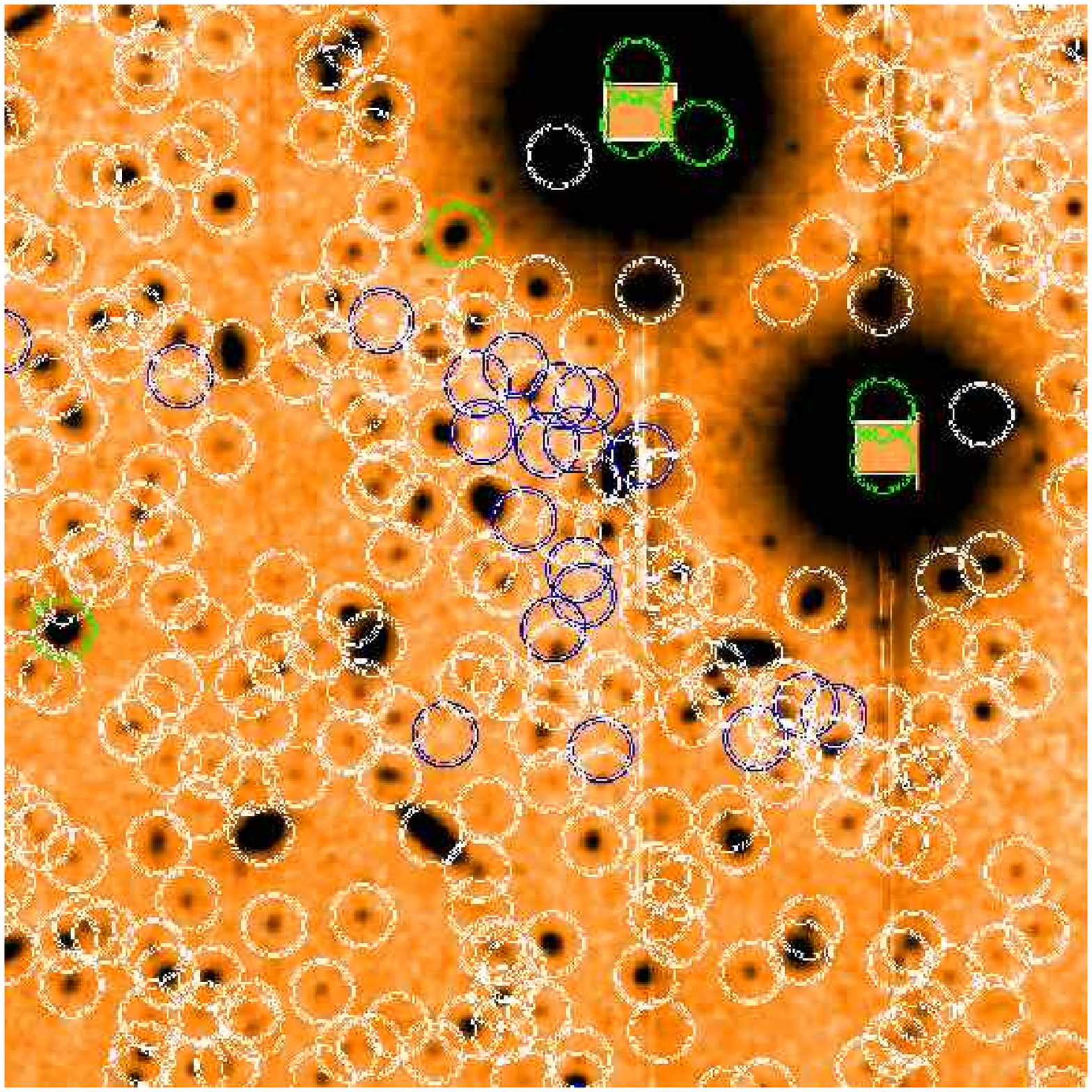}
  \includegraphics[angle=0,width=0.49\textwidth]{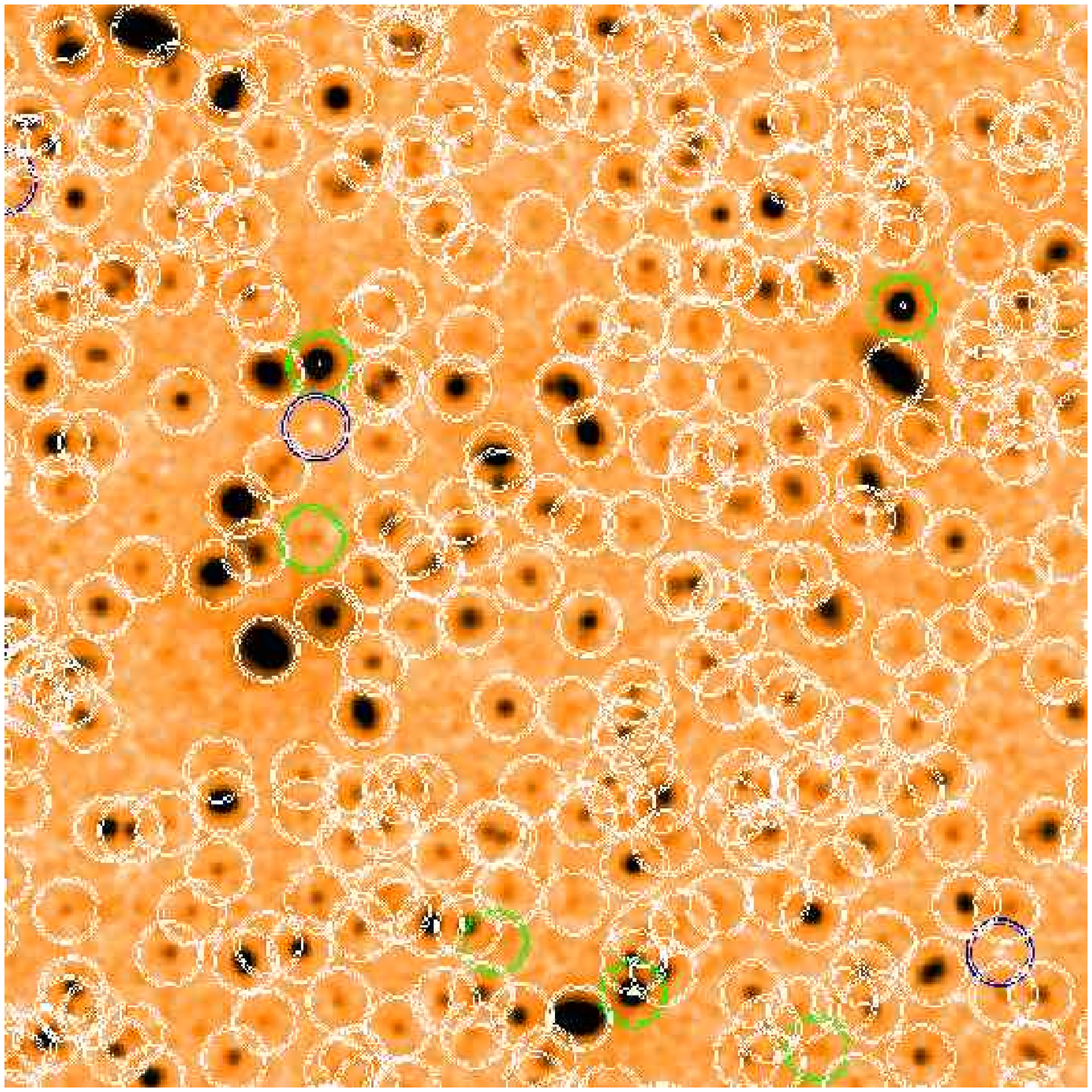}
\caption{ 1.5\arcmin \ x 1.5\arcmin Subaru i-band image. White circles
  mark the detected objects from the final catalogue, green circles
  mark objects eliminated from the final catalogues (saturated or
  corrupted magnitude, see text for details), whereas the blue circles
  mark false detections as detected on the negative image. It
  illustrates the fact, that there are many more detections in the
  negative image around bright objects (left panel) than in other
  areas of the image (right panel).
  \label{fals_detections}}
\end{figure*}
{We decided to use the ground based i-band for source detection
  mainly for two reasons: First, at the time when we started the
  photometric redshift determinations the HST coverage of our reduced
  H-band patches was not yet completed. Second, including space based
  data with their superb PSF would need a different technique (with
  HST images in various wavebands) to optimally construct the
  multi-waveband catalogue (see e.g. \citet{grazian:1} for the
  GOODS-MUSIC sample).}\\
We use a very similar approach as for the FDF \citep{fdf_data} in
constructing a multi-waveband catalogue in the COSMOS field.  First we
align by pixel all available filters (u$_{\rm CFHT}$, B$_{\rm
  Subaru}$, V$_{\rm Subaru}$, r$_{\rm Subaru}$, i$_{\rm Subaru}$,
z$_{\rm Subaru}$, Ks$_{\rm KPNO}$, afterwards referred to as u, B, V,
r, i, z, and K) to our H-band patches and derive i-selected
catalogues.  Please note that we had to compute a new astrometric
solution for the public K band, as the solution given in the images
was not sufficiently accurate. The program SCAMP \citep{scamp:1} was
applied for this procedure. Then we convolved all images to the same
seeing of 1.3\arcsec and ran SExtractor in the double image mode with
the detection threshold of $t=2.5$ and $n=3$ contiguous pixels (we
detect on the original i-band image with a seeing of 0.95\arcsec).
The 50 \% completeness limits as well as the seeing of the different
bands are listed in Table~\ref{tab:public_data}. The Table also
compares the depth of the different bands with the depth in the FDF.
Beside the u, z, and K bands, the COSMOS data set is roughly as deep as
the FDF. In the u-band, z-band as well as in the K band the FDF is
about 1 magnitude deeper.

In the FDF we did the source detection with $n=3$ contiguous pixels
and a threshold of $t=1.7$. This results in only a few false detections
(less than 1 percent). For the COSMOS data set we could not use the
same detection threshold as this would result in too many false
detections (measured on the negative image). We use a threshold of
$t=2.5$ in order to have false detections only at the 1 to 2 percent
level. Depending on the patch, the false detections as measured on the
negative image fluctuate between 0.6 percent and 2.6 percent.
Nevertheless this is not the real contamination rate, as a substantial
fraction of these false objects detected on the negative image are
distributed around bright objects. On the other hand the number
density on the positive side does \textit{not} increase around these
bright objects. Therefore it is most likely, that these false
detections are due to a reduction problem triggering false detections
only in the negative image. This can be seen in
Fig.~\ref{fals_detections} where we show a region around relative
bright objects together with the positive and negative detections.
Taking this into account, we conclude that we have a contamination
rate of about 1 percent or less in our catalogues.

In total we detected about 300~000 objects in the 12 patches (3
contiguous pixels and a detection threshold of 2.5). These patches
have a seeing of less then 1.3\arcsec (see above) in every band and
reside in a region which has at least 90\% of the maximum depth of
each band. Moreover we exclude all objects (based on the weighting
maps and SExtractor flags) with problems in the photometry (e.g. if
some pixels of an object are saturated or the magnitude is corrupted)
in at least one band in order to get a perfectly clean catalogue
suitable to derive photometric redshifts.  At this stage our clean
i-selected catalogue comprises 293~377 objects. As we work on the
individual patches and they slightly overlap, we analyse the objects
and found that about 99\% are unique entries.

A comparison between the i-band number counts in the COSMOS field and
in the literature as well as with the FDF are shown in
Fig.~\ref{nc_all_i}. There is a good agreement between the literature
and COSMOS number counts up to the limiting magnitude. Moreover, the
comparison with the FDF number counts shows an excellent agreement down
to the faintest bins. Only at the very bright end the FDF number
counts are slightly higher (most of the very bright objects are
saturated in the COSMOS i-band and thus not taken into account),
although not by more than 1$\sigma$ to $2\sigma$. The values of the
galaxy number counts can be found in Table~\ref{tab:nc}.

In order to avoid contamination from stars, we rely on two sources of
information: The star-galaxy classifier of the detection software
SExtractor, and the goodness of fit of the photometric redshift code.
We first exclude all bright ($i < 22.5^m$) starlike objects
(SExtractor star galaxy classifier $>0.97$).  Then we exclude all
fainter objects whose best fitting stellar spectral energy
distribution (SED) -- according to the photometric redshift code --
gives a better match to the flux in the different passbands than any
galaxy template ($2\ \chi_{star}^2 < \chi_{galaxy}^2$). These objects
are subsequently flagged as stars and removed from our galaxy
catalogue. In order to test the accuracy of our procedure we further
inspect by eye (in one patch) on the public ACS data if the objects
flagged as stars are extended. It turned out that about half of these
objects are extended in the ACS data and only half are point-like.
Although this approach \citep[see][]{gabasch:1} of excluding stars
works very well in the FDF where the seeing of the detection image is
only 0.55\arcsec, it is much less effective in the COSMOS i band with
a seeing of 0.95\arcsec.

In total 4803 (1.6 \%) objects were classified as stars and removed
from our sample (see also Fig.~\ref{histo_galaxies} for the
photometric redshift distribution of all starlike objects excluded
from the galaxy catalogue). Please note that most of the bright
point-like objects are saturated in one of the bands and already
removed in the first cleaning step (see above). Therefore our final
i-band selected galaxy catalogue comprises 288~574 objects.

\section{Spectroscopic redshifts}\label{sec:spec}

\begin{figure}
  \centering
  \includegraphics[angle=0,width=0.50\textwidth]{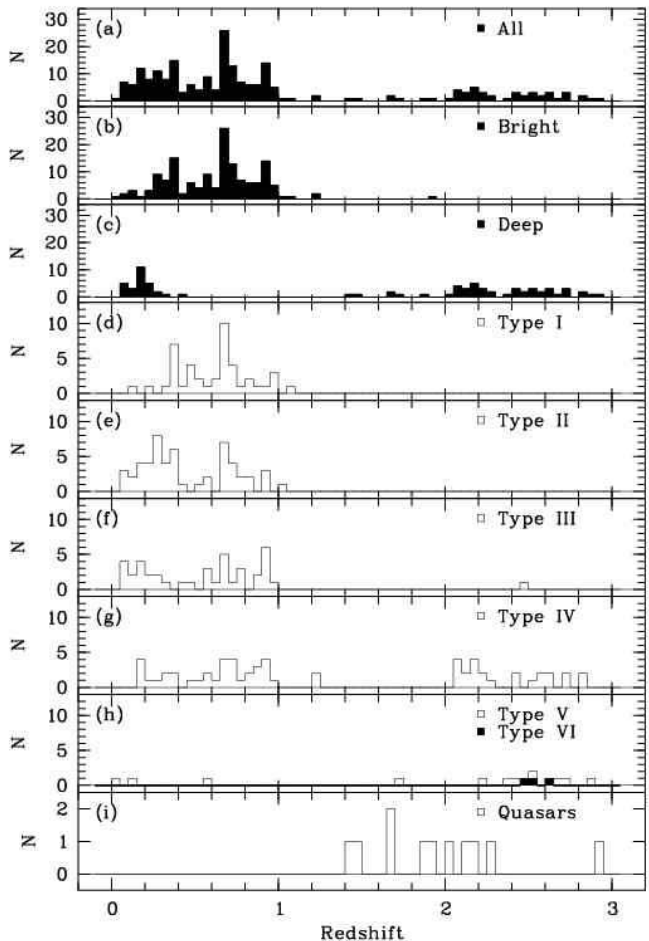}
\caption[]{Redshift distribution of the COSMOS spectroscopic sample for (a) 
  all extragalactic objects with trustworthy redshifts, (b) the
  corresponding objects of the bright-galaxies mask, (c) the
  corresponding objects of the zCOSMOS-deep mask, (d) -- (h) galaxies
  of different object type, and (i) the quasars. The types range from
  I = ellipticals (or passively evolving) to V = extreme starbursts
  \citep[see ][]{NOL04}. In diagram (h) galaxies of type V
  (white bars) as well as type VI (strong Ly$\alpha$ emitters, black
  bars) are plotted. The redshift resolution is $\Delta z = 0.05$.}
\label{fig_tz}
\end{figure}

Photometric redshifts need to be calibrated by spectroscopic redshifts
for objects covering a wide range of galaxy types and redshifts.
Hence, we selected spectroscopic data of the COSMOS field
\citep{scoville:1} from the ESO archive taken with the VIMOS
spectrograph \citep{LeF03} at the ESO VLT in the context of the
zCOSMOS project
\citep{lilly:4}. zCOSMOS is a Large ESO Programme with 600\,hr of
observing time aiming at the characterisation of the distribution and
properties of galaxies out to redshifts of $z \sim 3$. The project is
divided into two parts.  zCOSMOS-bright focuses on relatively bright
galaxies with $I_{\rm AB} < 22.5$ at redshifts $0.1 < z < 1.2$. In
order to include such galaxies in our spectroscopic control sample we
selected the VIMOS mask `zCOSMOS\_4-4' with 312 slits centred at
10h00m28.20s +02$^{\rm o}$15$'$45.0$''$. This mask was observed
$1.5$\,h with the red low-resolution LR\_red grism ($R \sim 210$),
covering a wavelength range from 5500 to 9500\,\AA{}.  zCOSMOS-deep
aims at galaxies with $1.5 < z < 2.5$. For this
\citet{lilly:4} selected targets using the $BzK$ criteria of
\citet{DAD04} and the $UGR$ `BX' and `BM' selection of \citet{STE04}.
We obtained a representative sample of such galaxies, reducing the
VIMOS mask `zCOSMOS\_55\_faint' with 239 slits at 10h00m27.67s
+02$^{\rm o}$10$'$23.0$''$, observed $4.5$\,h with the blue
low-resolution LR\_blue grism ($R \sim 180$), covering a wavelength
range from 3700 to 6700\,\AA{}.

The spectra of both VIMOS masks were reduced using VIPGI
\citep{SCD05}. This software package is designed to reduce the
multi-object VIMOS spectra in a quite automatic way. The reduction was
performed using essentially standard methods. The jittered sequences
of individual exposures (five different positions) allowed to
efficiently correct for fringing and to obtain a high-quality sky
subtraction. The one-dimensional composite spectra were extracted by
means of the S/N-optimised \citet{HOR86} algorithm. The final
spectra are flux calibrated and corrected for atmospheric absorption
bands.

Redshifts were derived using the cross-correlation based algorithm
described by \citet{NOL04}. In order to obtain the redshift and a
rough spectral type, we used a sequence of six empirical templates
\citep[see ][{for more details}]{NOL04} essentially differing
in their UV-to-optical flux ratio, but also showing different
strengths of nebular emission lines. For the comparison to photometric
redshifts we consider galaxies with trustworthy ($> 90$\% confidence)
spectroscopic redshifts only. Due to low S/N (particular at high
redshift) and the limited wavelength ranges of the spectra certain
redshifts could only be derived for about half of the objects,
i.e. 272 of 551 spectra. For the zCOSMOS-deep mask the success rate
(33\%) is significantly lower than for the bright-galaxy mask (62\%).
The total sample of 272 objects with trustworthy redshifts includes 49
stars and 12 quasars.  Hence, the final sample of galaxies comprises
211 objects. The majority of them (147 or 70\%) belongs to the
bright-galaxy sample. On the other hand, the subsample of 36 galaxies
with redshifts $1.5 < z < 3$ only consists of objects from the
zCOSMOS-deep sample. These high-redshift galaxies (quasars excluded)
correspond to 45\% of the objects identified in the deep sample.
Fig.~\ref{fig_tz} shows the redshift distribution of the full
spectroscopic control sample, the `zCOSMOS\_4-4' mask, the
`zCOSMOS\_55\_faint' mask, the different galaxy types defined by
\citet[][I = ellipticals, V = extreme starbursts, VI = strong
Ly$\alpha$ emitters]{NOL04}, and the quasars.

After cross-correlating (by visual inspection) the 211 spectroscopic
objects with the final i-band selected catalogue in the 12 patches we
end up with a final sample of 162 spectroscopic redshifts used to
calibrate the photometric redshifts. {As the spectroscopic
  redshift sample was merely used to calibrate the photometric
  redshift and to measure the accuracy of the photometric redshift
  code, we decided to use only objects with an unique cross
  identification. Therefore we excluded all objects where the
  cross-matching between the photometric and spectroscopic catalogue
  was not unique, where the objects of the photometric catalogue show
  saturation in one (or more) image and where the spectroscopic
  distance determination was not very reliable (e.g. low S/N ratio).}

\section{Photometric redshifts}
\label{sec:photoz}

\begin{figure*}
\centering
  \includegraphics[angle=0,width=0.49\textwidth]{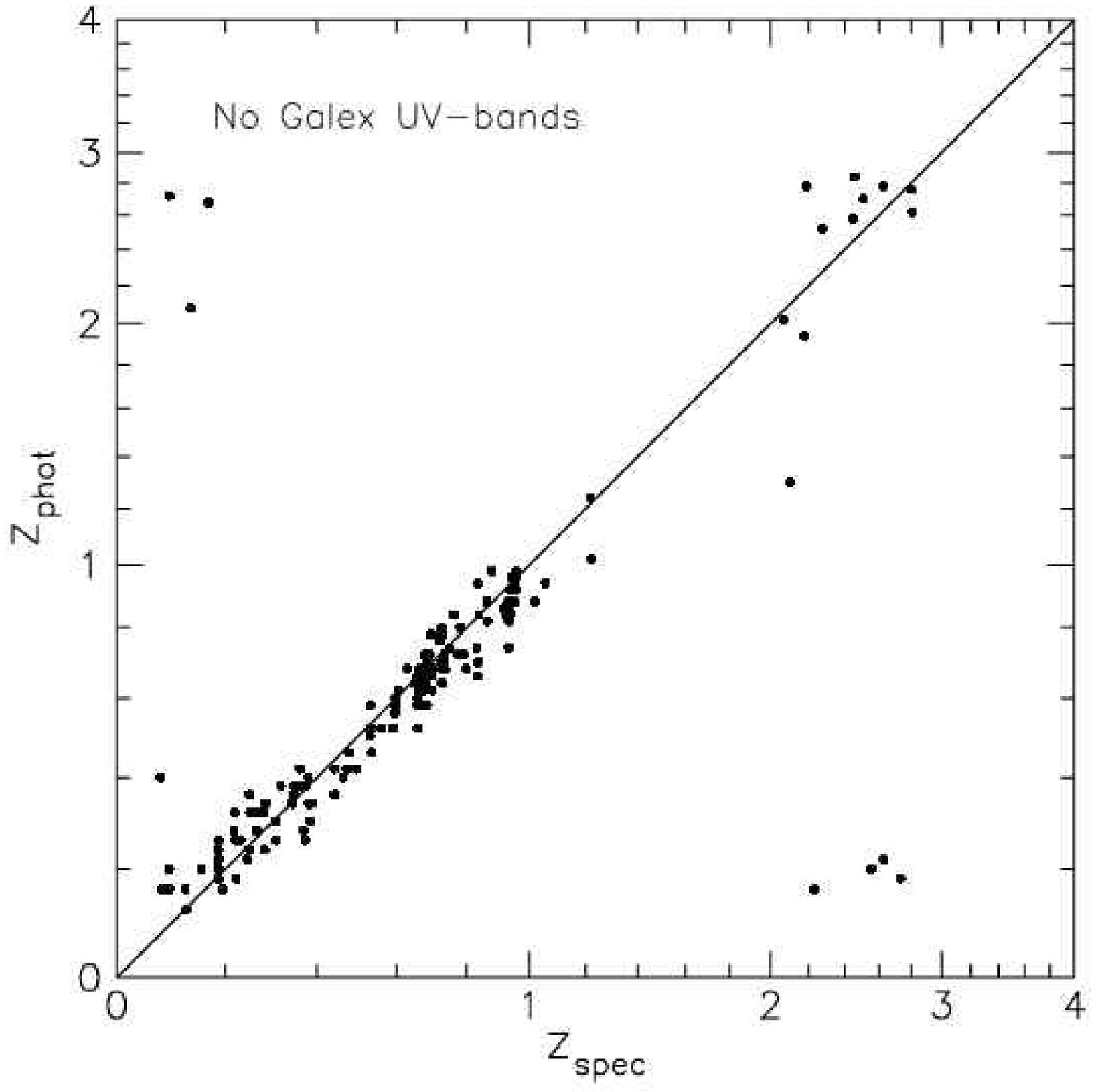}
  \includegraphics[angle=0,width=0.49\textwidth]{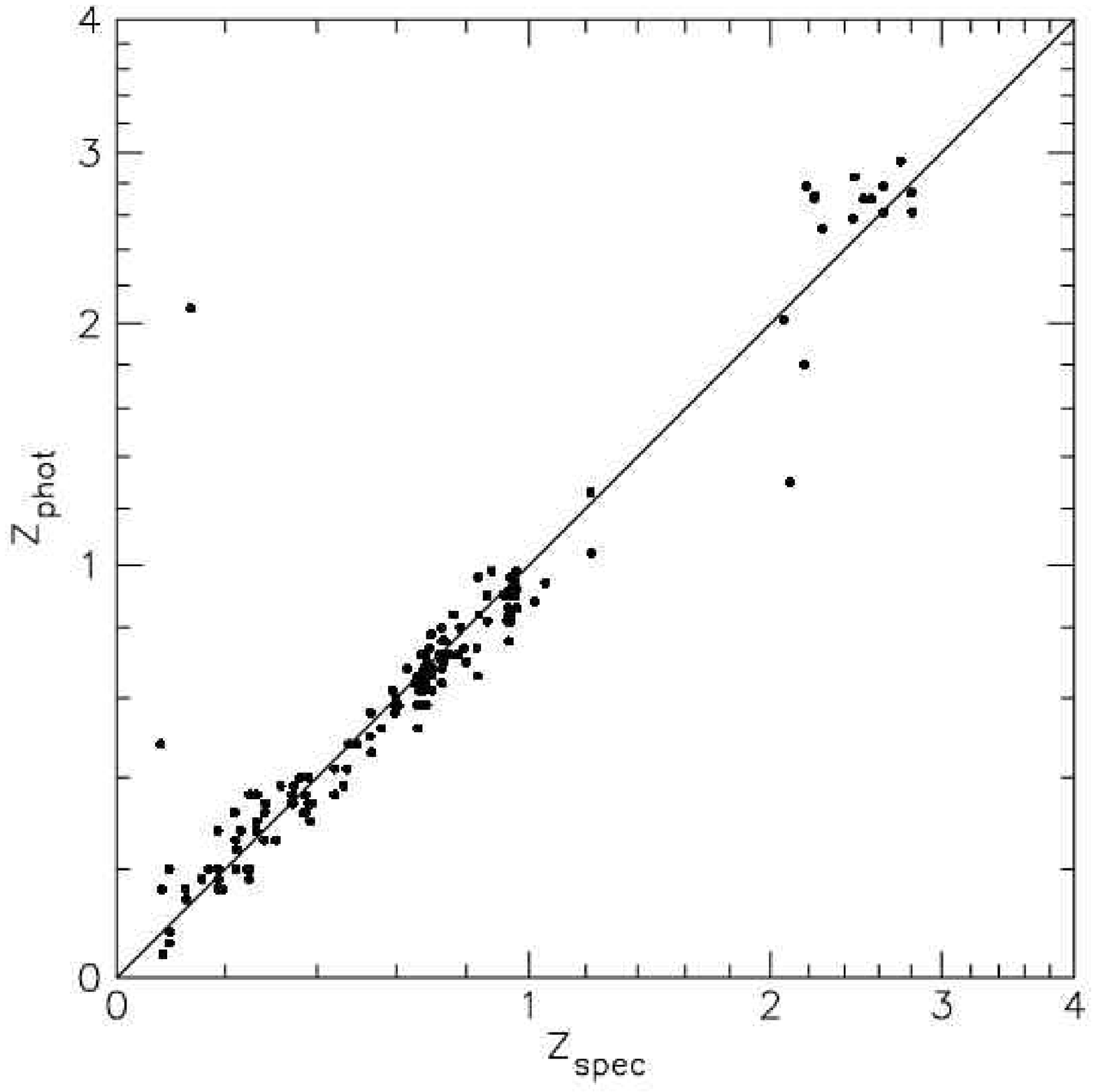}
\caption{Comparison of spectroscopic and
  photometric redshifts in the COSMOS field (162 galaxies). Left
  panel: u band to K band are used to derive the photometric redshifts.
  Right panel: u band to K band as well as the GALEX FUV and NUV bands
  are used to derive the photometric redshifts.
  \label{comp_photoz_spec}}
\end{figure*}

\begin{figure*}
\centering
  \includegraphics[angle=0,width=0.49\textwidth]{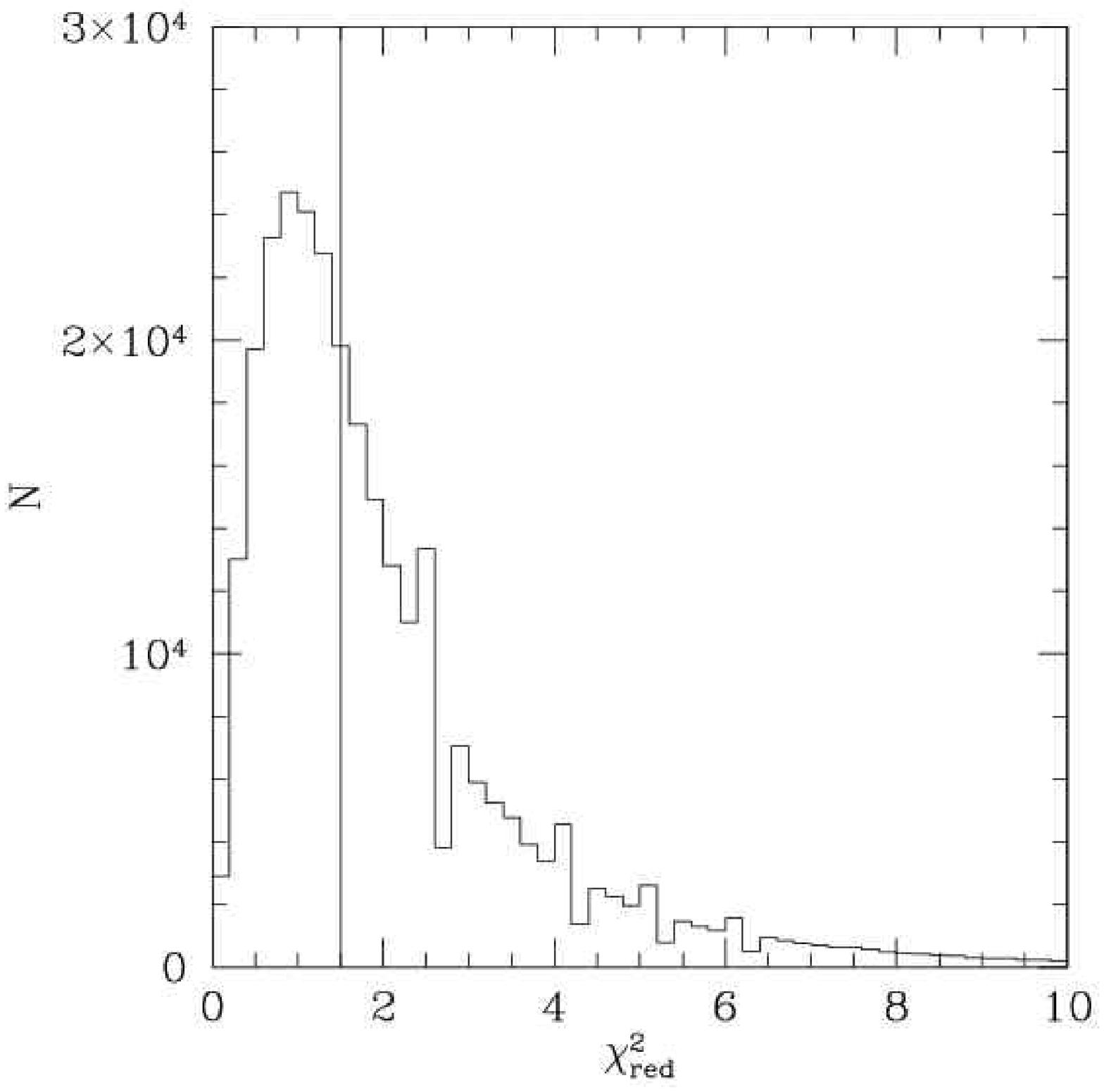}
  \includegraphics[angle=0,width=0.49\textwidth]{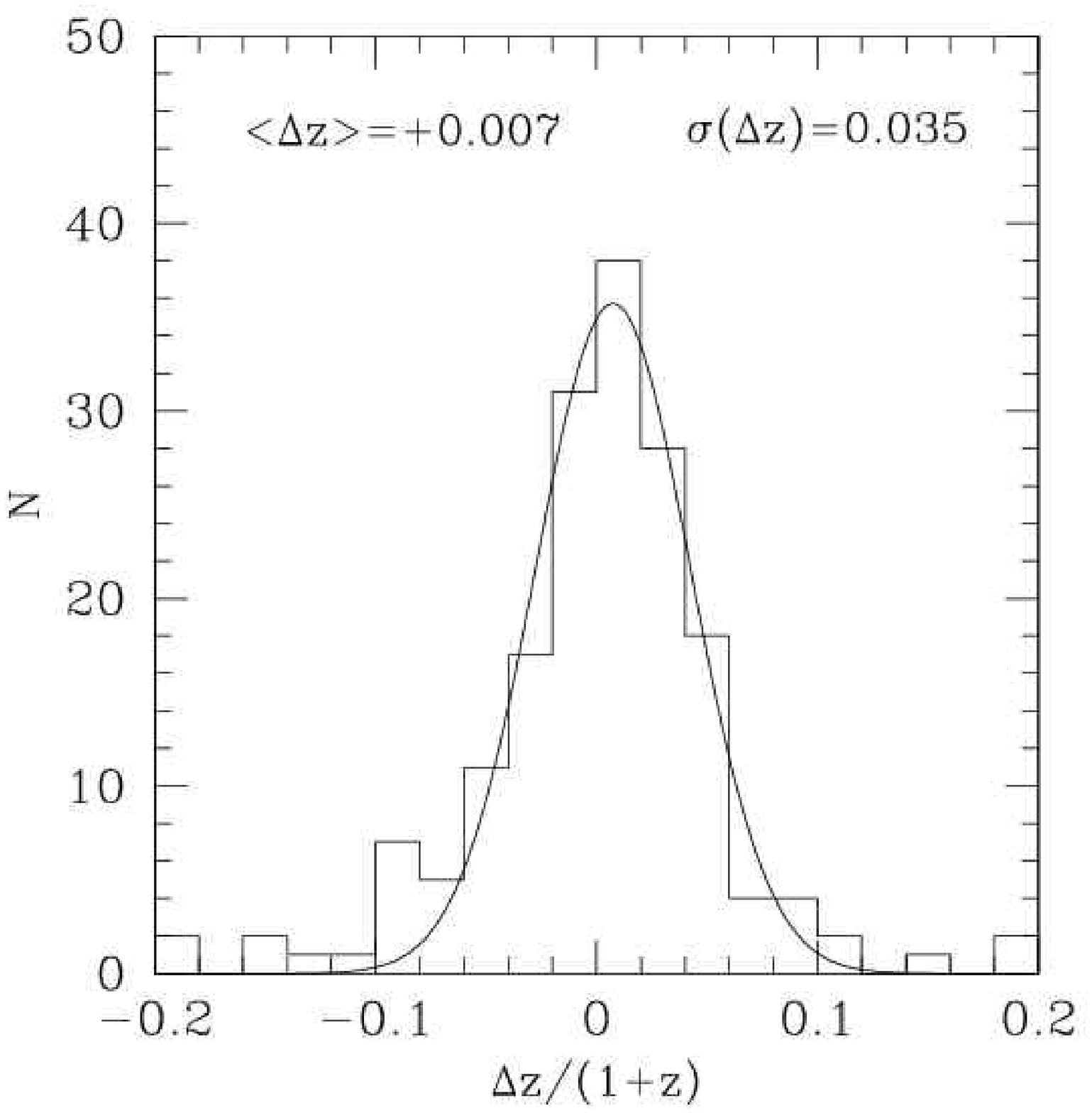}
\caption{ Left panel: Histogram of the reduced $\chi^2$ for all galaxies in the
COSMOS field as obtained for the best fitting
template and redshift. The dotted vertical line indicates the median
reduced  $\chi^2$ of 1.5. Right panel: Histogram of the photometric 
redshift errors. The error distribution can be approximated by a Gaussian centred at
0.007 with an rms of 0.035 (solid line).  
  \label{histo_photoz}}
\end{figure*}

\begin{figure*}
\centering
  \includegraphics[angle=0,width=0.49\textwidth]{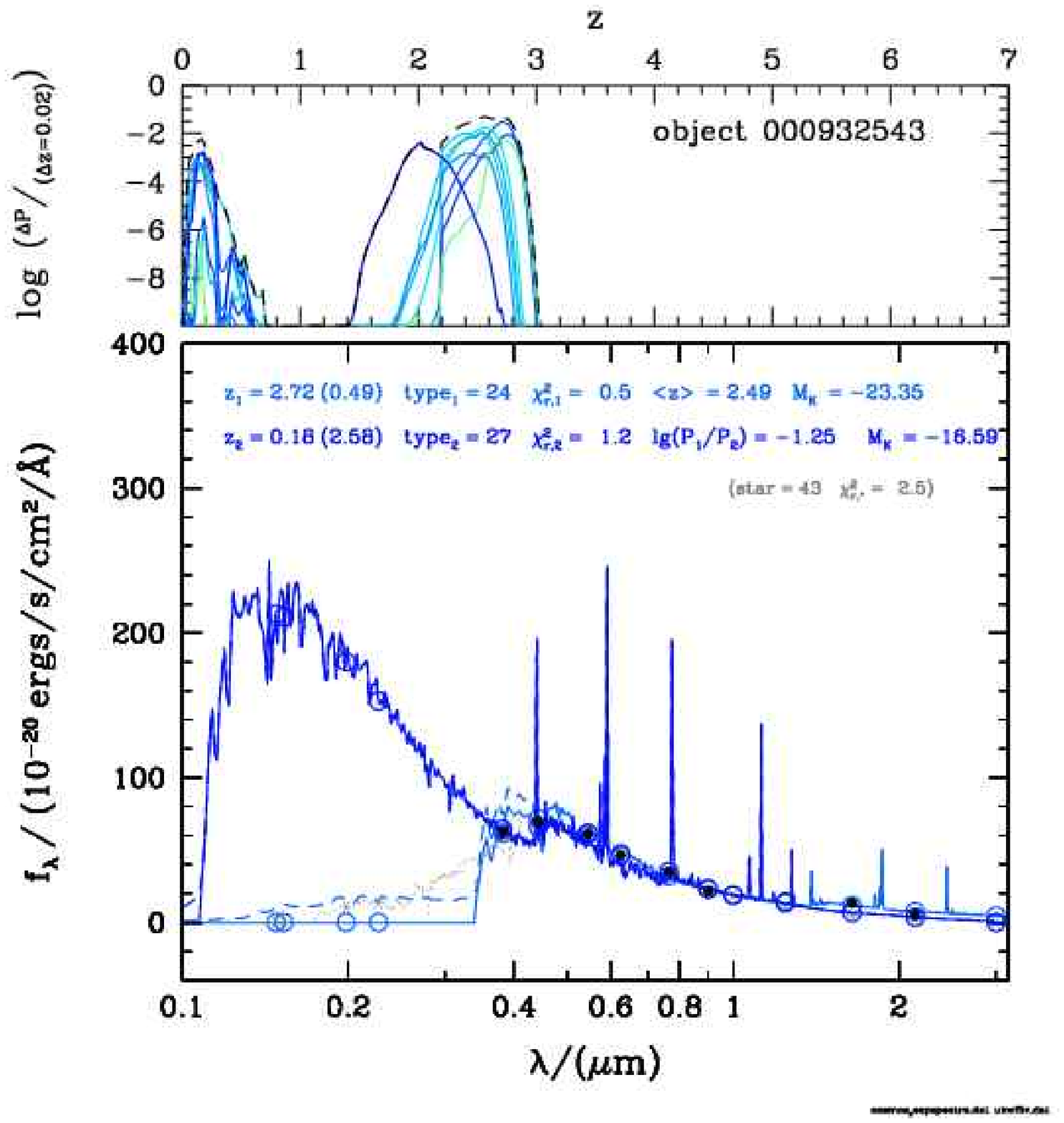}
  \includegraphics[angle=0,width=0.49\textwidth]{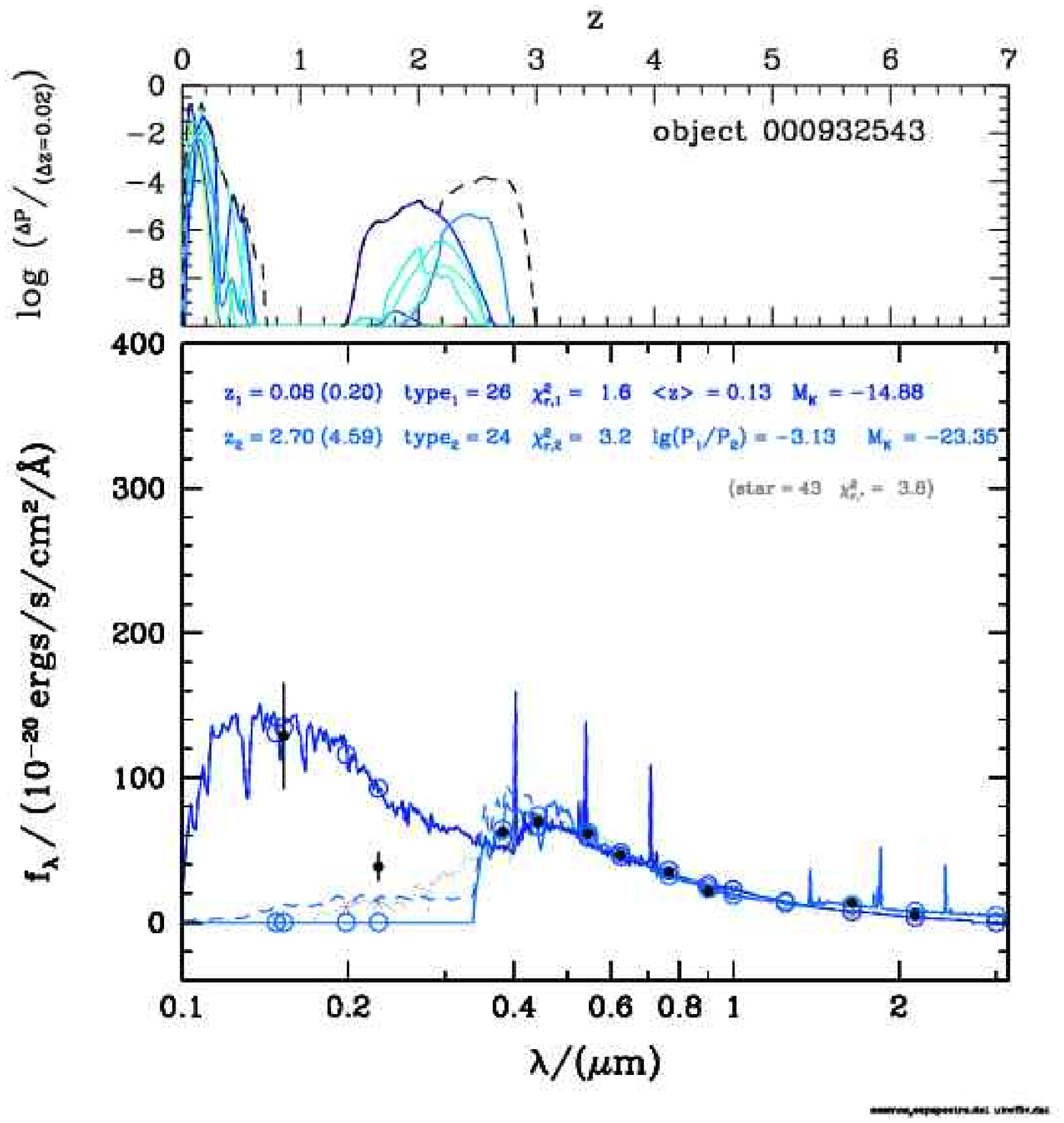}
\caption{ Redshift probability distribution (upper part) and photometric redshift
  fit (lower part) of the spectroscopic object 000932543 with
  $z_{spec}=0.093$.  The different SEDs are colour coded. The open
  circles (lower part) represent the SED integrated within the various
  filter transmission curves, whereas the black dots represent the
  measured fluxes. The redshift for the best fitting SED ($z_1$) as
  well as for the second best fit ($z_2$) are also given. Left panel:
  Only 8 bands (u-band to K-band) are used to derive the photometric
  redshift yielding a wrong $z_{phot}$ of 2.72.  Right panel: 10 bands
  (FUV, NUV, u-band to K-band) are used to derive the photometric
  redshift yielding a $z_{phot}$ of 0.08 very close to the
  spectroscopic redshift.
  \label{photoz_sedfits}}
\end{figure*}

\begin{figure*}
\centering
  \includegraphics[angle=0,width=0.49\textwidth]{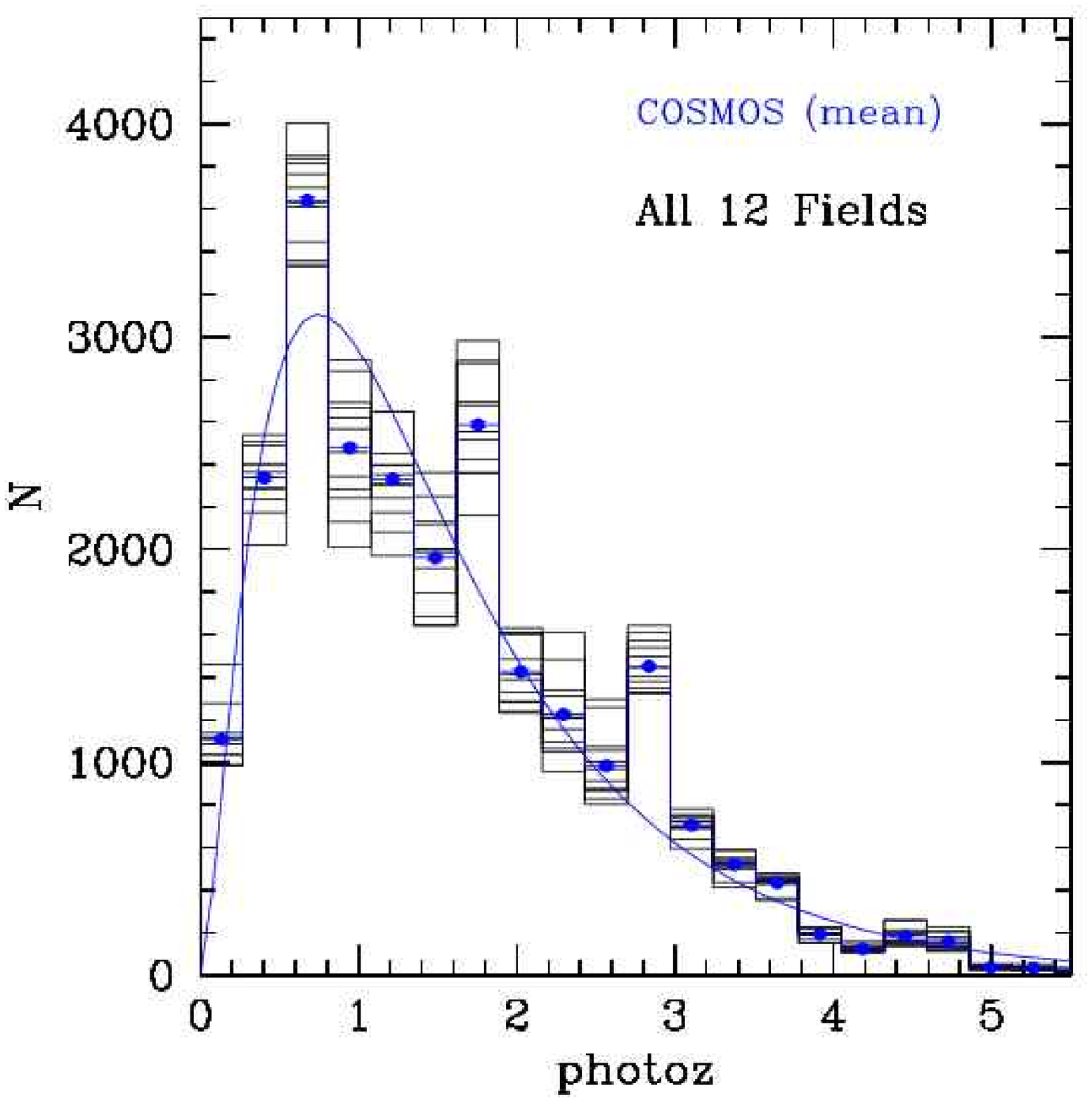}
  \includegraphics[angle=0,width=0.49\textwidth]{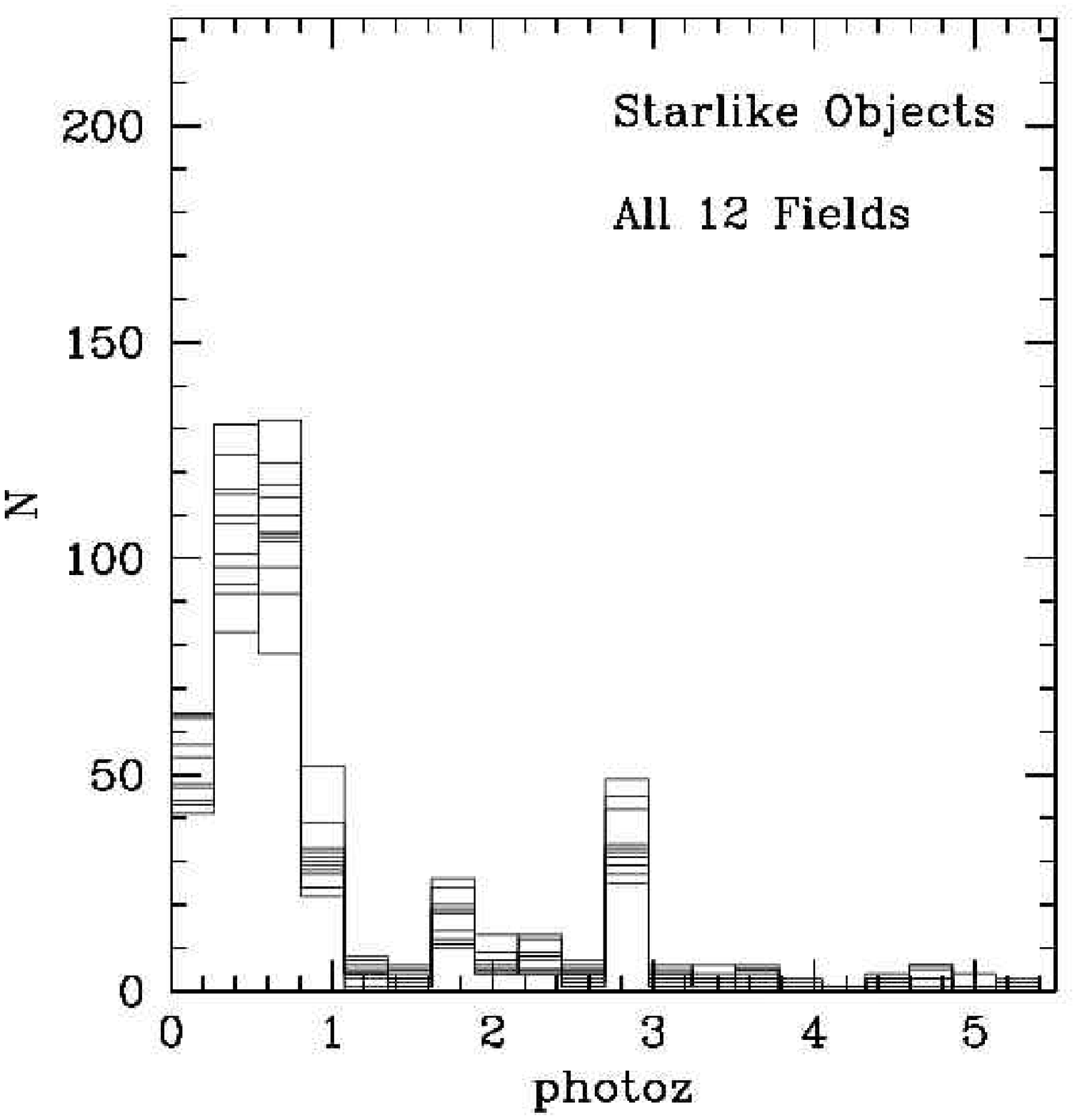}
\caption{ Left panel: Redshift
  number distribution of all galaxies in the 12 patches. The mean
  number distribution is shown as blue dots. The blue solid line
  represents a fit to the mean galaxy distribution using the approach
  of \citet{brainerd:1} (see text for details). Right panel: Redshift
  distribution of objects classified as stars (see text for details)
  and eliminated from the final galaxy catalogues. Please note the
  different scaling by a factor of 20 in the left and right panel.
  \label{histo_galaxies}}
\end{figure*}

A summary of the photometric redshift technique used to derive the
distances to the galaxies can be found in \citet{bender:1} and
\citet{gabasch:1}. Before deriving the photometric redshifts we
checked and fine-tuned the calibration of our photometric zeropoints by
means of colour-colour plots of stars. We compared the colours of
stars with the colours of stellar templates from the library of
\citet{pickles:1} converted to the COSMOS filter system.  In general,
corrections to the photometric zeropoints of only a few hundredth of a
magnitude were needed to obtain an good match to the stars and best
results for the photometric redshifts (if compared to the
spectroscopic ones).  Only in the u-band and in the K-band, the
correction were in the order of a few tenths of a magnitudes.  A
comparison between the reduced KPNO K-band and our K-band (both
convolved to the same seeing of 1.3\arcsec) in patch 01c showed, that
although the total magnitudes agreed very well, the fixed aperture
magnitudes (especially of relatively faint sources) differ
systematically by a few tenths of a magnitude.  Therefore we decided to
correct the KPNO zeropoint by matching the faint sources in the KPNO
image with those in our own field (in the fixed aperture used to
derive the photometric redshifts). Moreover when calculating the
photometric redshifts, we artificially increased the magnitude errors
in the K-band by $0.25^m$ (added in quadrature to the SExtractor
errors) to reduce the relative weight of this slightly problematic
band. Therefore, we rely mostly on the accurate photometry of the NIR
H band.

In order to avoid contamination from close-by objects, we derived
object fluxes for a fixed aperture of $2.0\arcsec$ ($1.5 \times$
seeing) from images which had been convolved to the same point spread
function (PSF; $1.3$\arcsec). A redshift probability function P(z) was
then determined for each object by matching the object's fluxes to a
set of 29 template spectra redshifted between $z=0$ and $z=10$ and
covering a wide range of ages and star-formation histories.

In Fig.~\ref{comp_photoz_spec} (left panel) we compare 162 high
quality galaxy spectroscopic redshifts with the photometric redshifts.
Although there is a good agreement in the redshift range between
$z\sim0.2$ and $z\sim1.2$, it is clear from
Fig.~\ref{comp_photoz_spec}, that there is a degeneracy between high
redshift ($z\sim2.5$) and low redshift ($z\sim0.2$) objects (10
catastrophic outliers with \mbox{$\Delta z / (z_{spec}+1) \gsim
  0.2$}). This degeneracy stems from the relatively red u-band. In
Fig.~\ref{photoz_sedfits} we show the redshift probability function as
well as the SED fits to the observed flux of the spectroscopic object
000932543. Although the spectroscopic redshift is $z_{spec}=0.093$,
the best fitting photometric redshift is $z_{phot}=2.72$. On the other
hand, there is also a low redshift peak around $z_{phot}=0.18$ in the
redshift probability function (but with a lower probability). Moreover
Fig.~\ref{photoz_sedfits} also shows that both, the high redshift as
well as the low redshift solutions are hard to disentangle as long as
no information in the UV is available (as they differ mainly in the
UV). Therefore we decided to include in the determination of the
photometric redshifts also the GALEX FUV and NUV
bands\footnote{The data were taken from:\\
  http://irsa.ipac.caltech.edu/data/COSMOS/}.

As we do not want to convolve all the images to a seeing of 5\arcsec
(GALEX PSF), we decided to use another approach to include the UV
fluxes in our photometric redshift estimation. Similar to the optical
and NIR bands we used a fixed aperture of $\sim 1.5 \times$ PSF, i.e.
7.5\arcsec.  As there were no obvious features in the colour-colour
plots of stars including the UV bands, we could not fine-tune the
calibration of the zeropoints by means of colour-colour plots of
stars.  Therefore we optimised the zeropoints by using the SED fits of
our galaxies with very good photometric redshifts (if compared with
the spectroscopic ones). Please note that this approach can not derive
accurate UV fluxes, but gives only a very rough flux estimation in the
two UV bands. Nevertheless, it is now possible to break the degeneracy
between the high redshift and low redshift solution. This can be best
seen in Fig.~\ref{photoz_sedfits} where we show one of the
catastrophic outliers. Only by including the NUV and FUV fluxes (right
panel) we are able to obtain a photometric redshift of $z=0.08$, hence
very close to the spectroscopic redshift of $z_{spec}=0.093$.  This
approach drastically reduces the number of our catastrophic
photometric redshift outliers (see Fig.~\ref{comp_photoz_spec}).

In Fig.~\ref{comp_photoz_spec} (right panel) we compare the final
photometric and spectroscopic redshifts of the 162 galaxies. The
agreement is very good and we have only 3 catastrophic outliers. The
right panel of Fig.~\ref{histo_photoz} shows the distribution of the
redshift errors.  It is nearly Gaussian and scatters around zero with
an rms error of \mbox{$\Delta z / (z_{spec}+1) \approx 0.035 $}.
Fig.~\ref{histo_photoz} (left panel) presents the $\chi^2$
distribution of the best fitting templates and photometric redshifts
for all the objects. The median value of the reduced $\chi^2$ is 1.5
and demonstrates that the galaxy templates describe the vast majority
of galaxies very well.  {Please note that although the
  photometric redshift accuracy of a single object does not
  considerably improve by adding our H-band to the publically
  available data, the number of catastrophic outliers decreases by
  nearly a factor of 2.  Moreover, the H-band was also very useful to
  find and
  address the problems in the public K-band as mentioned before.}\\
The galaxy redshift histogram of all objects in the different patches
is shown in Fig.~\ref{histo_galaxies}. The mean galaxy distribution
can be very well described by Equation~(\ref{eqn:brainerd}) introduced
by \citet{brainerd:1},
\begin{equation}
p_z(z) = Const \times  \frac{\beta z^2}{\Gamma (3/\beta) z_0^3} \exp
\left(-(z/z_0)^{\beta}\right)   
\label{eqn:brainerd}
\end{equation}
where $Const$, $z_0=<z>\frac{\Gamma (3/\beta)}{\Gamma (4/\beta)}$ and
$\beta$ are free parameters with $<z>$ being the first moment of the
distribution and $\Gamma$ the Gamma function. The best fitting values
are: $Const=6206$, $z_0=0.107$, and $\beta=0.611$.

Please note that if we analyse the galaxy photometric redshift
histogram with a binning of $\Delta z=0.1$ there are three clearly
visible peaks below redshift of $z=2$: one at $z_{phot}=[0.6, 0.7]$,
one at $z_{phot}=[0.9, 1.0]$, and one at $z_{phot}=[1.7, 1.8]$.
Interestingly, we also find peaks in the spectroscopic redshift
histogram (see also Fig.~\ref{fig_tz}) with at least 10 galaxies at
$z_{spec}=[0.657, 0.669]$, $z_{spec}=[0.672, 0.683]$, and
$z_{spec}=[0.926, 0.941]$.

\section{UV luminosity function and density}
\label{sec:uvlf}

\begin{figure}
  \centering
  \includegraphics[angle=0,width=0.50\textwidth]{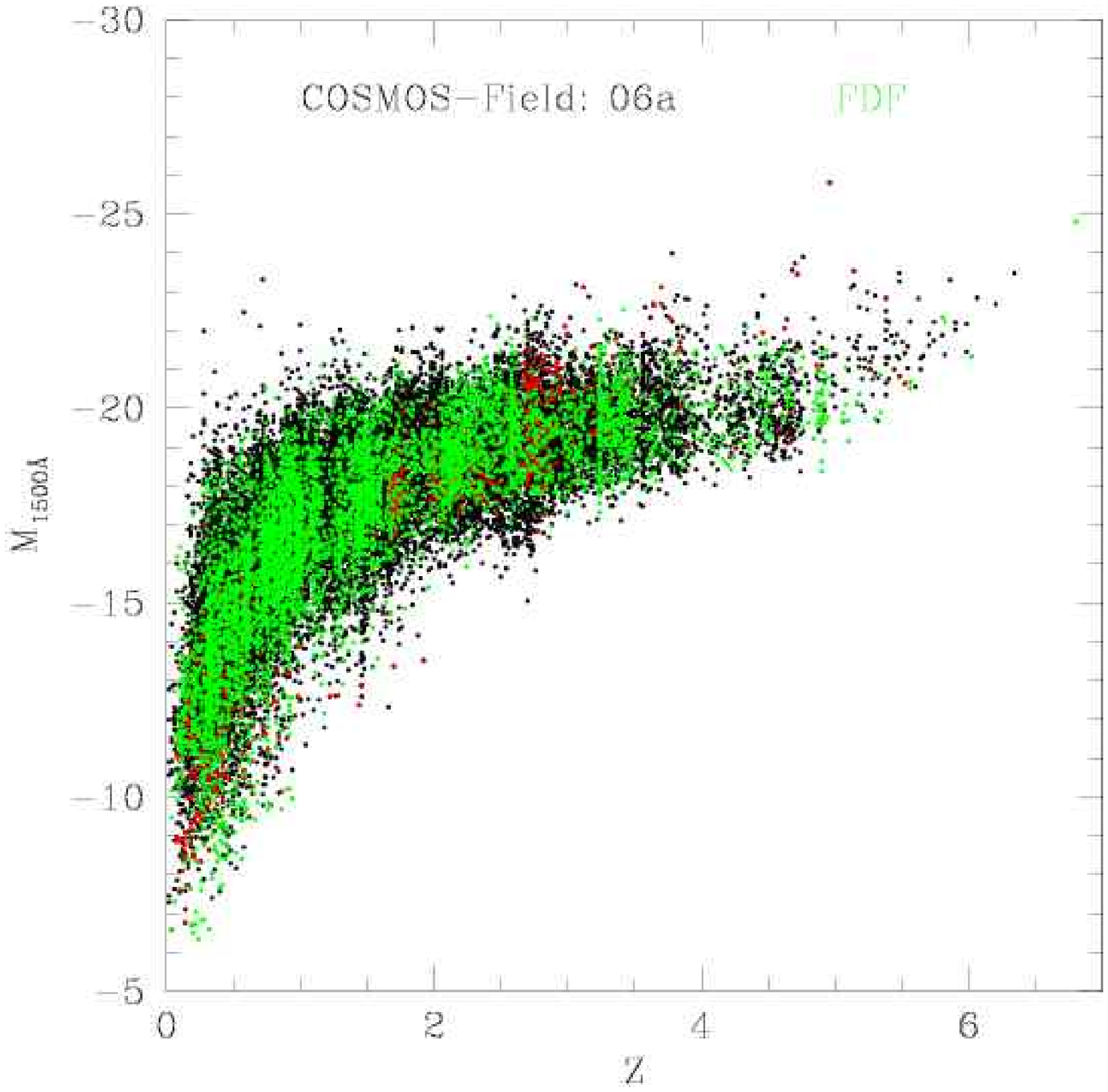}
\caption[]{Absolute UV-band magnitudes of galaxies in the COSMOS 06a
  patch (black dots) as a function of redshift. The green dots are
  derived in the FDF whereas the red dots represent COSMOS objects
  classified as stars if a more conservative criterion in separating
  stars from galaxies ($\chi_{star}^2 <\chi_{galaxy}^2 \Rightarrow$
  star; see text for details) is applied.
\label{fig:absmag_1500}}
\end{figure}

\begin{figure*}
  \centering
  \includegraphics[angle=0,width=0.32\textwidth]{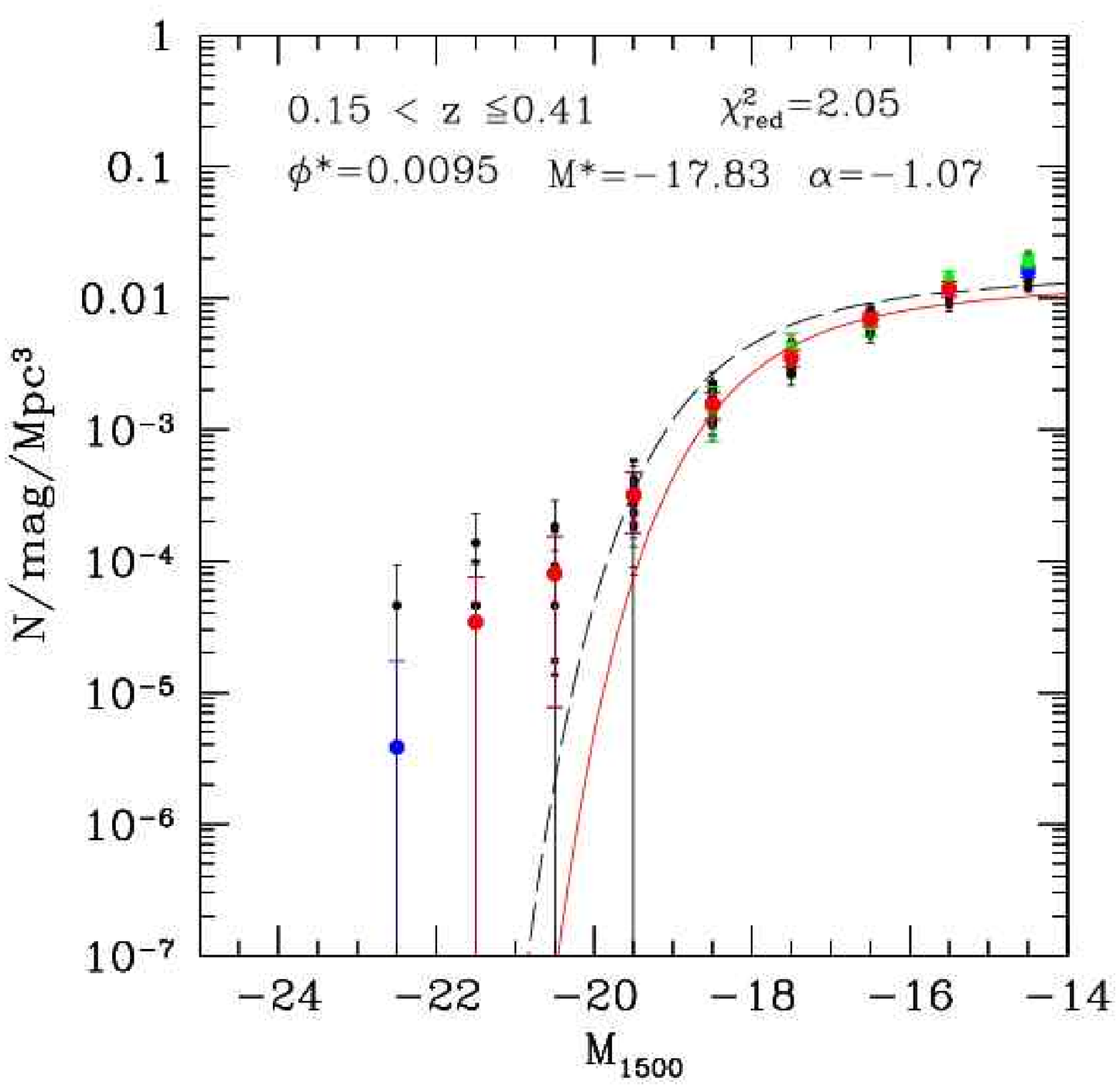}
  \includegraphics[angle=0,width=0.32\textwidth]{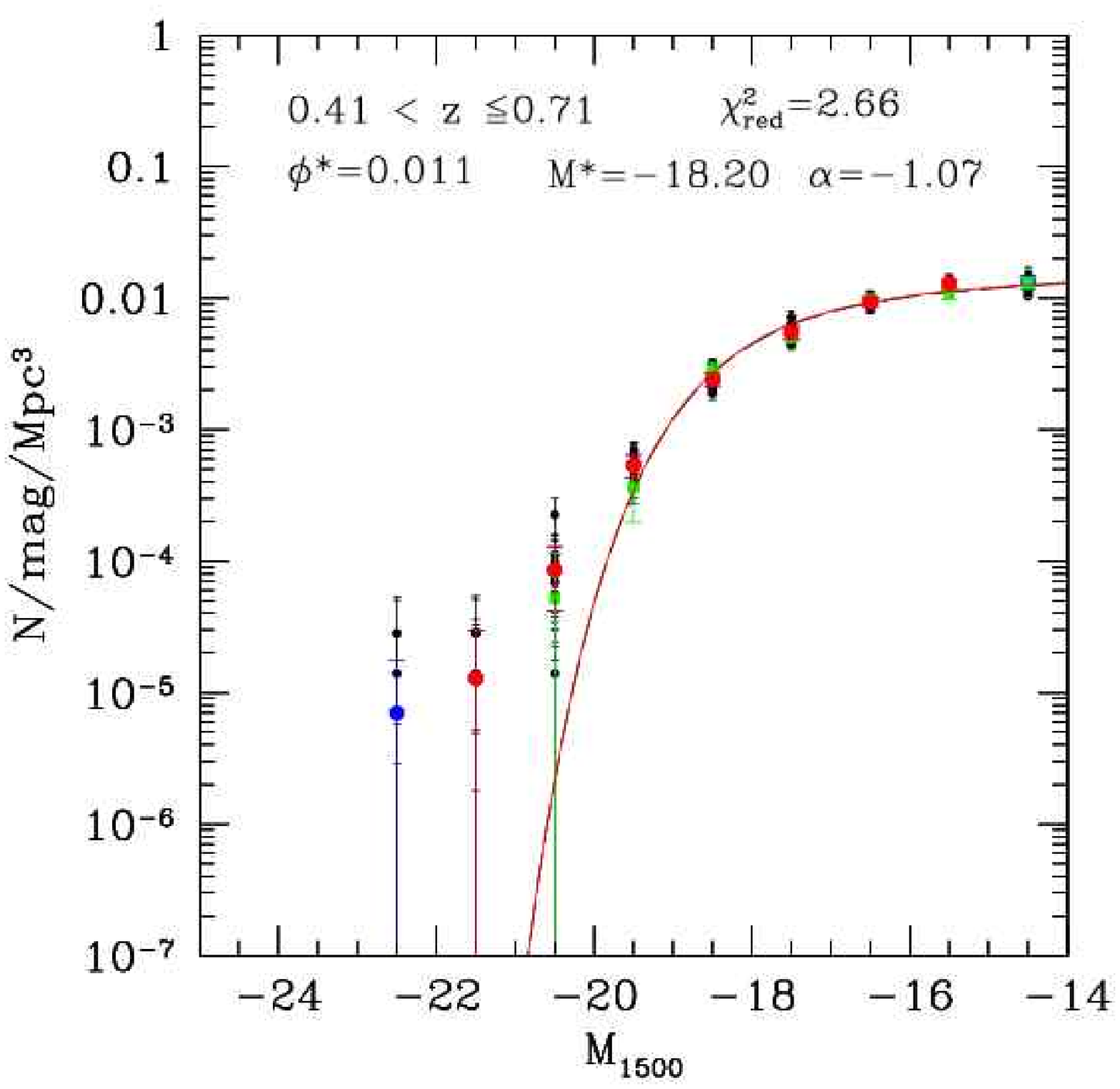}
  \includegraphics[angle=0,width=0.32\textwidth]{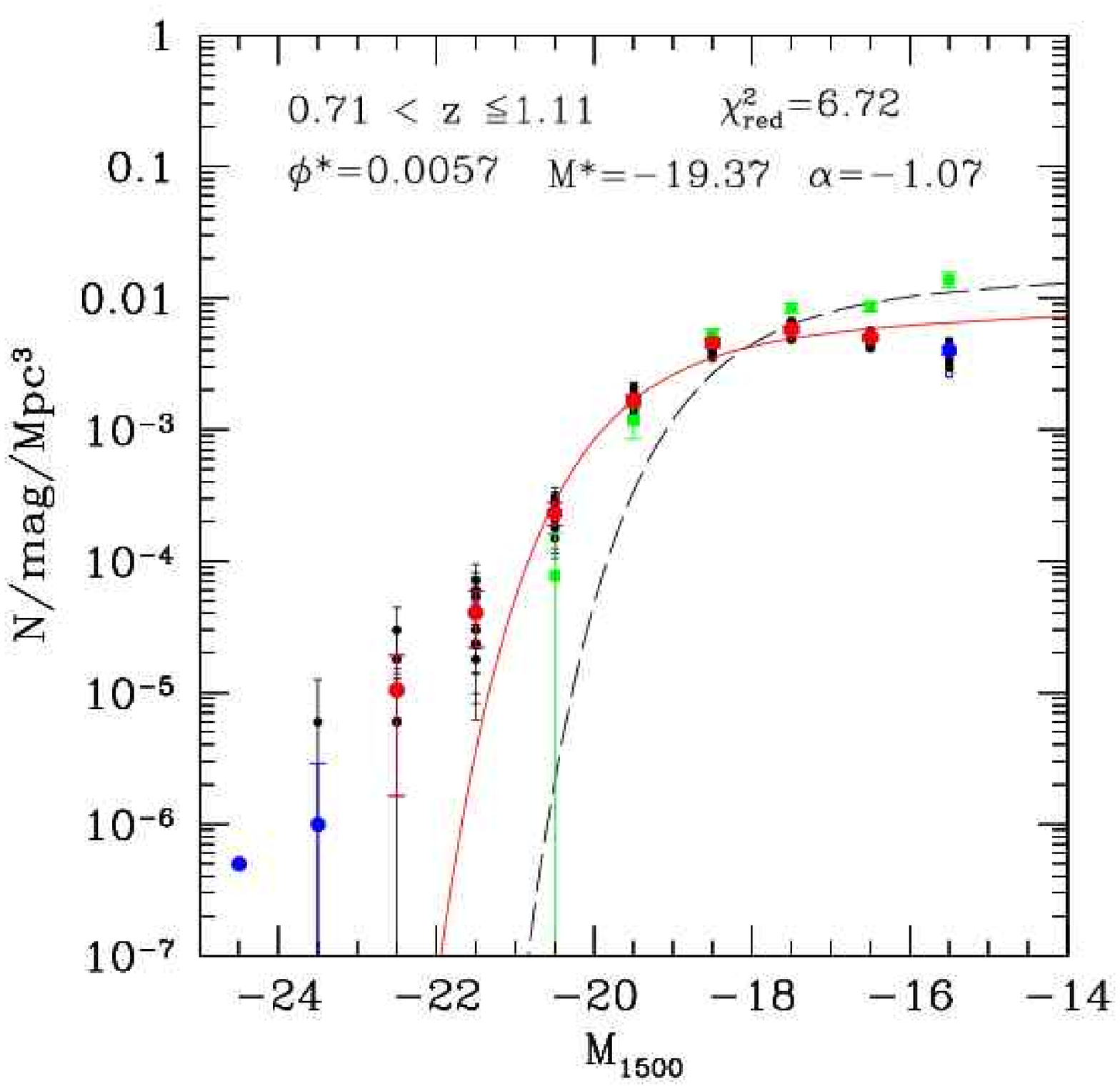}
  \includegraphics[angle=0,width=0.32\textwidth]{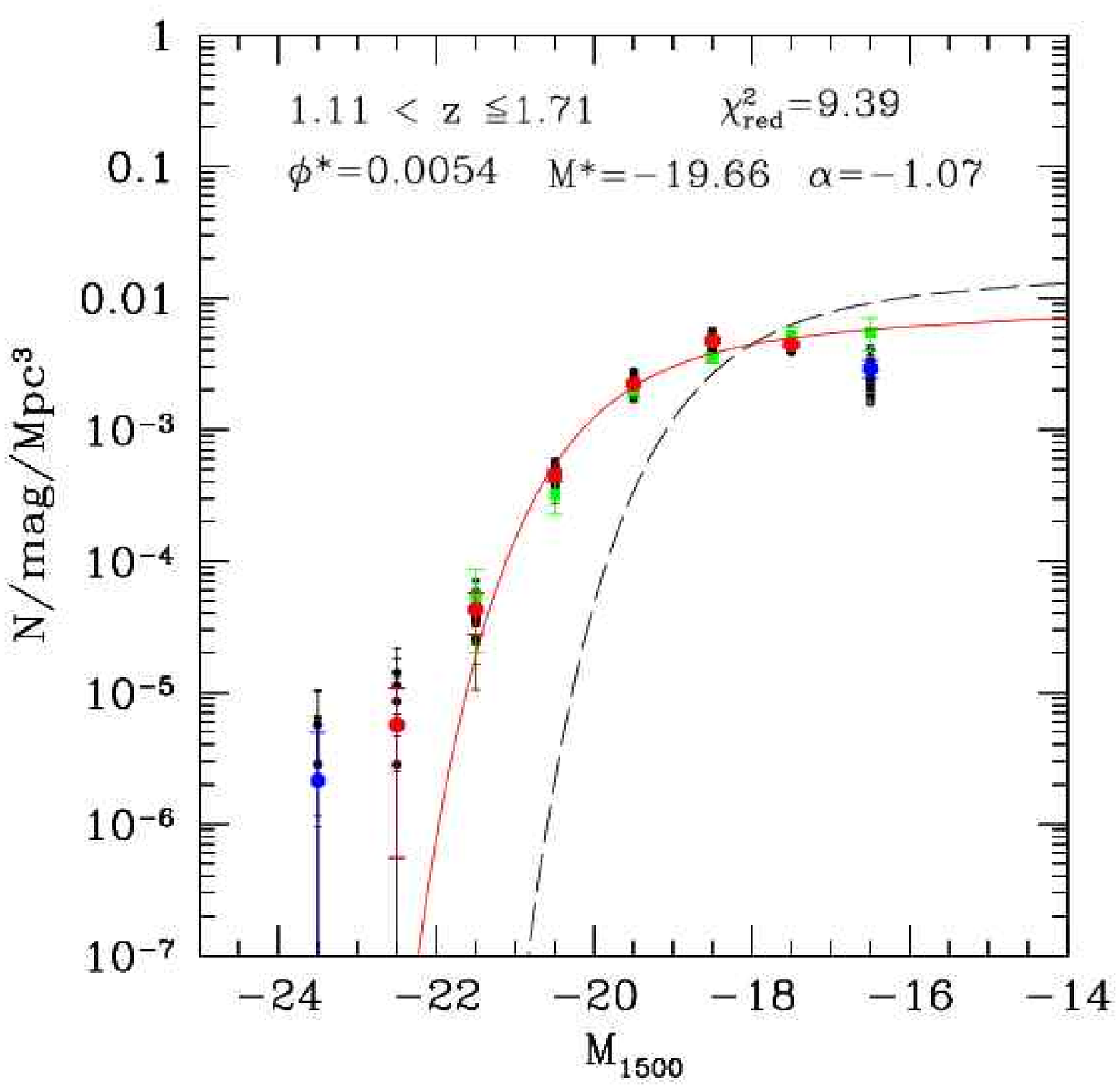}
  \includegraphics[angle=0,width=0.32\textwidth]{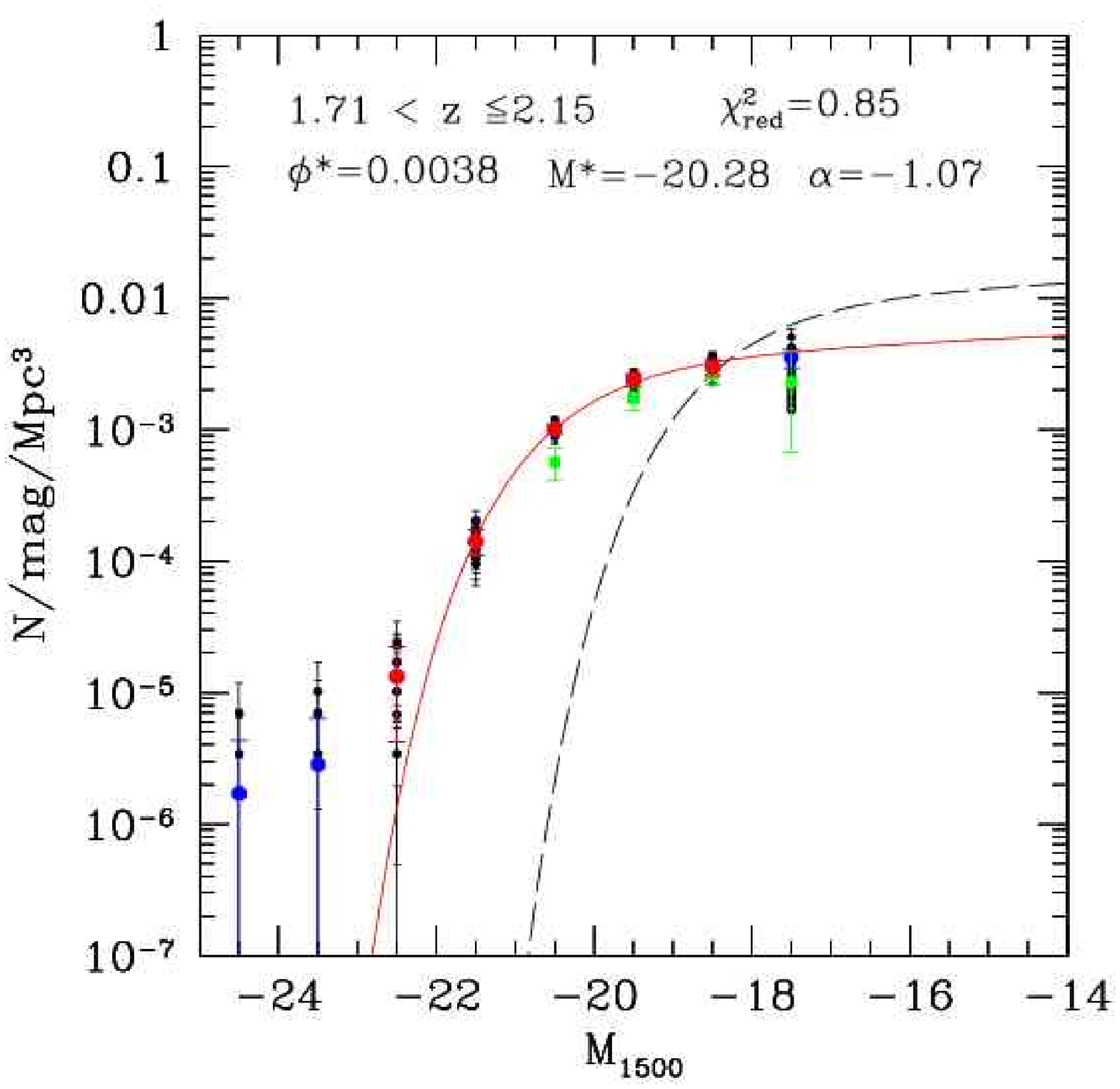}
  \includegraphics[angle=0,width=0.32\textwidth]{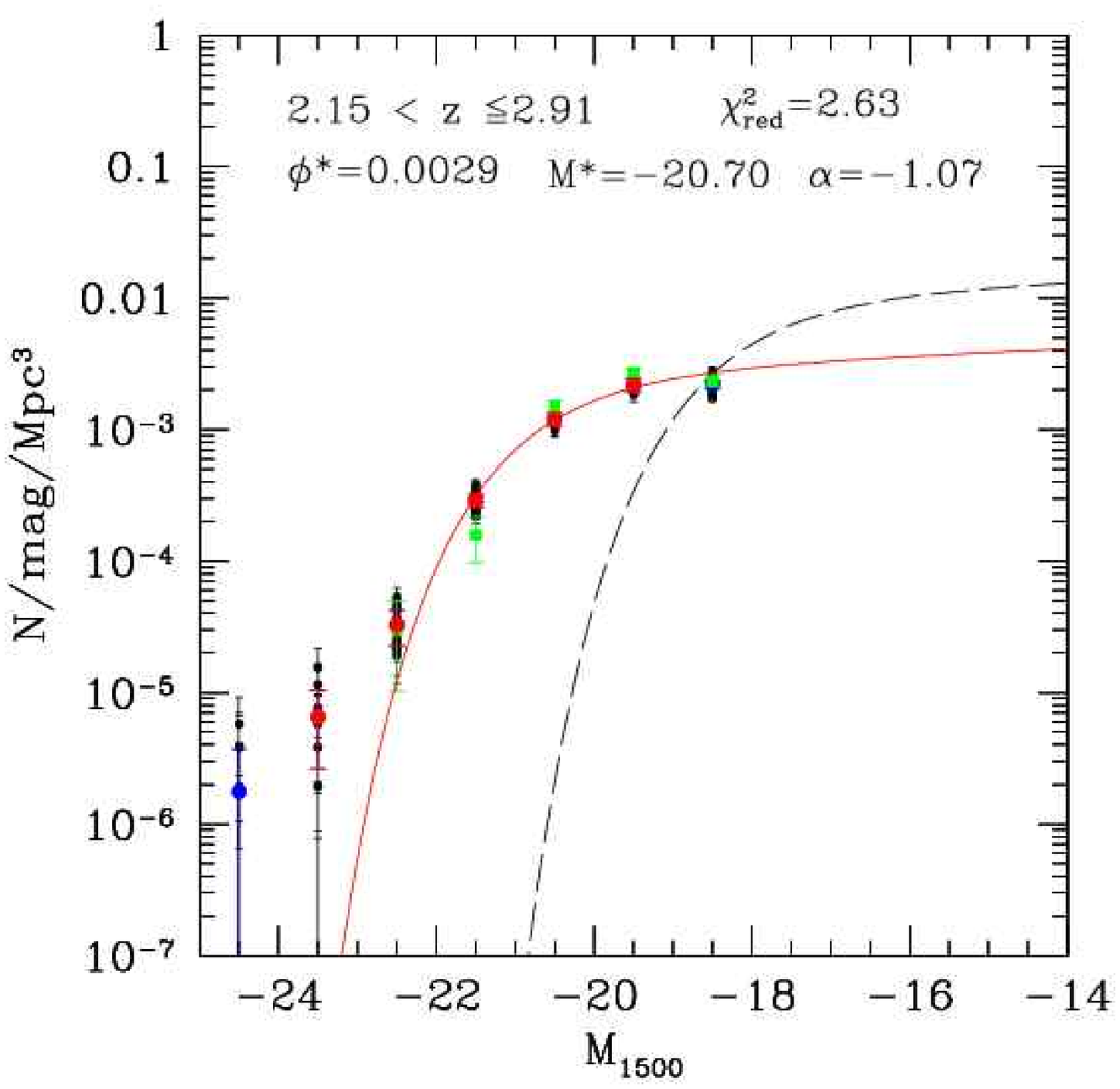}
  \includegraphics[angle=0,width=0.32\textwidth]{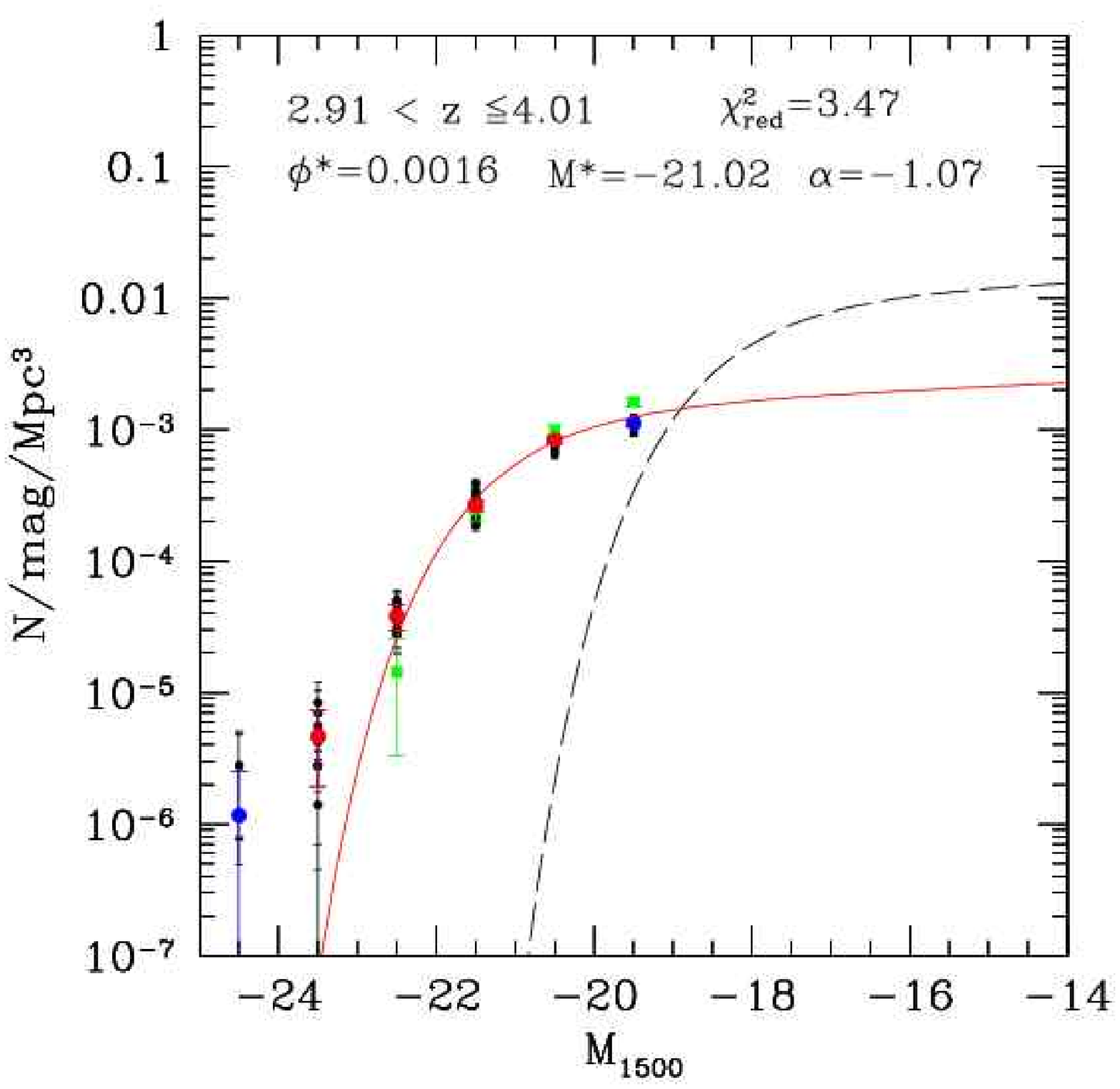}
  \includegraphics[angle=0,width=0.32\textwidth]{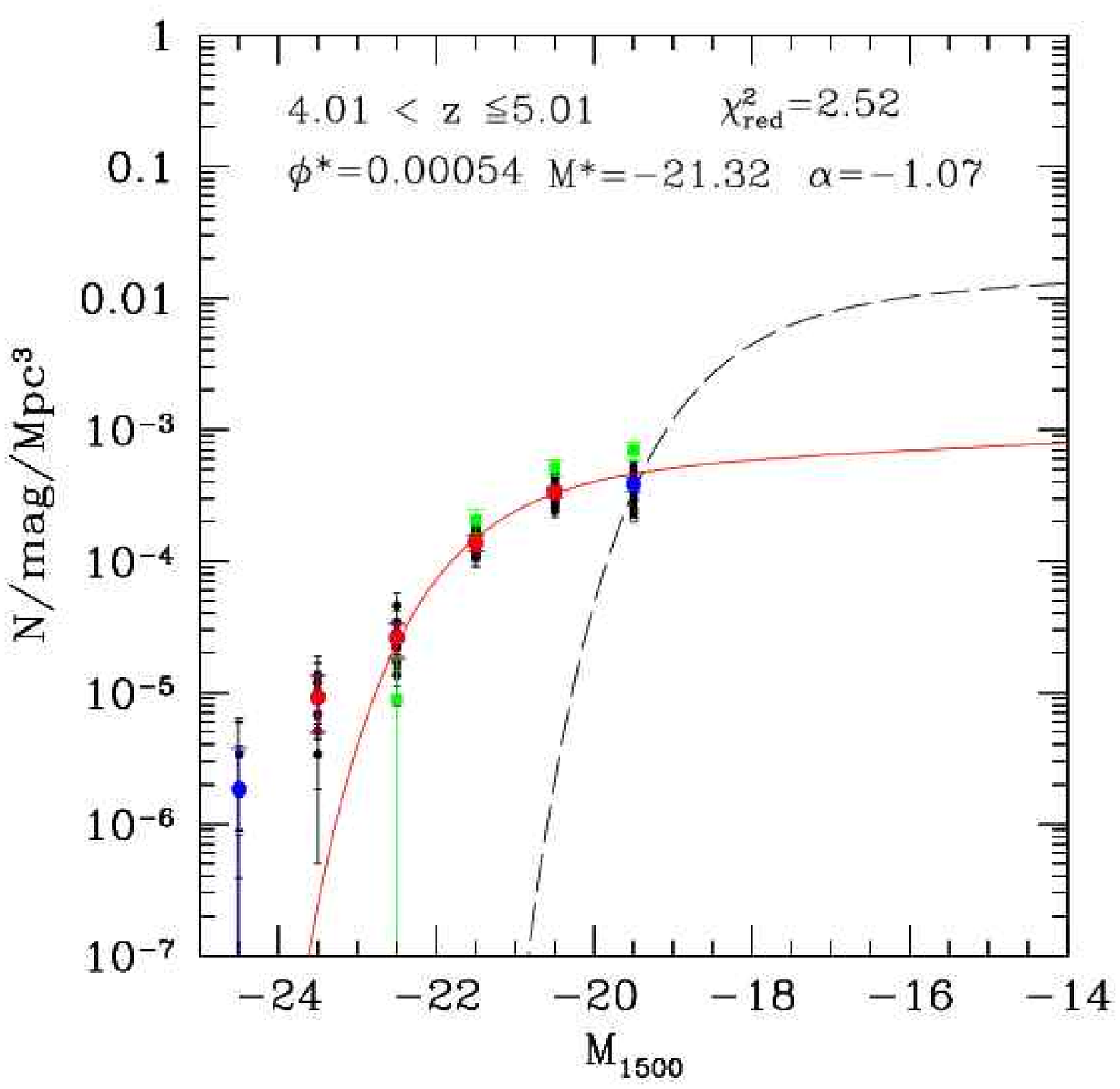}
  \includegraphics[angle=0,width=0.32\textwidth]{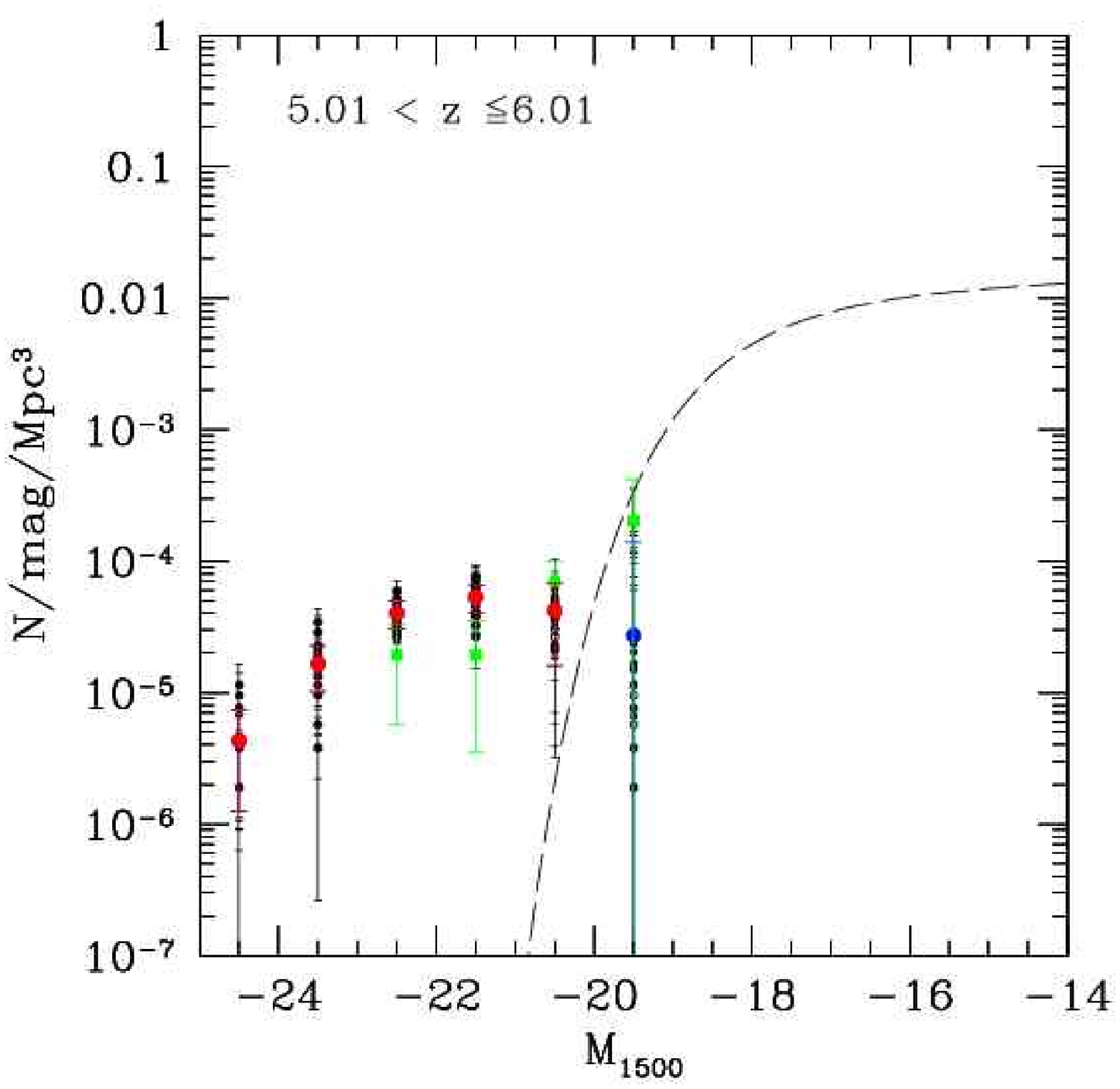}
\caption{ 
  Luminosity functions at \textit{1500~\AA } from low redshift
  (\mbox{$\langle z\rangle=0.3$}, upper left panel) to high redshift
  (\mbox{$\langle z\rangle=5.5$}, lower right panel). The filled
  (open) black symbols show the luminosity function corrected
  (uncorrected) for $V/V_{max}$ in the various patches. The red and
  the blue dots represent the mean LF in the field. The green squares
  represent the LFs as derived in the FDF.  The fitted Schechter
  functions for a fixed slope $\alpha$ are shown as solid red lines
  (only the red dots are used to derive the best fitting Schechter
  functions). Note that we only fit the luminosity functions to
  $\langle z\rangle=4.5$. The parameters of the Schechter functions
  can be found in Table~\ref{tab:schechter_fit_1500}. The Schechter
  fit for redshift $\langle z\rangle=0.6$ is indicated as a dashed
  black line in all panels.\label{fig:uvlf}}
\end{figure*}

\begin{figure*}
\centering
  \includegraphics[angle=0,width=0.49\textwidth]{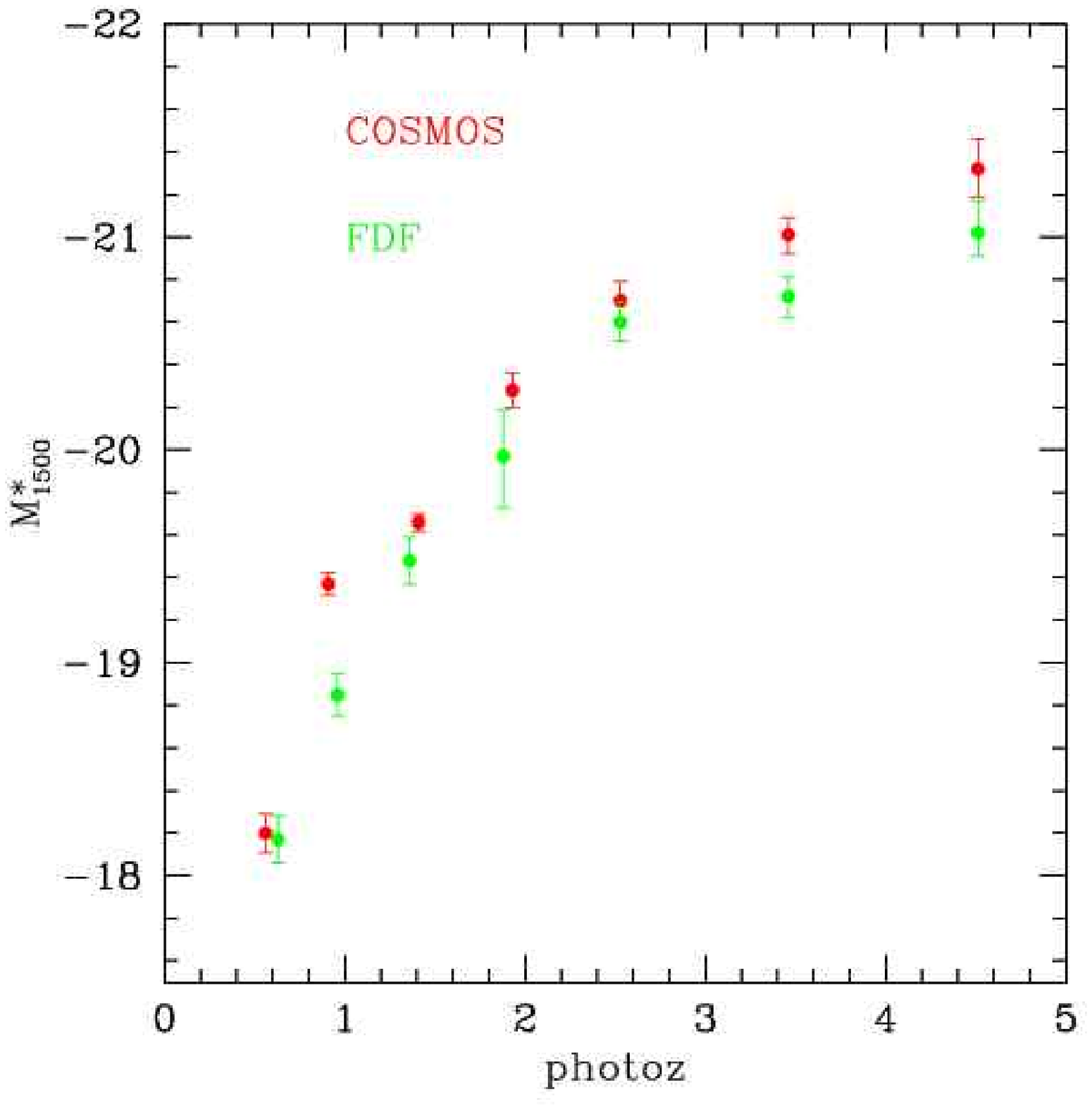}
  \includegraphics[angle=0,width=0.49\textwidth]{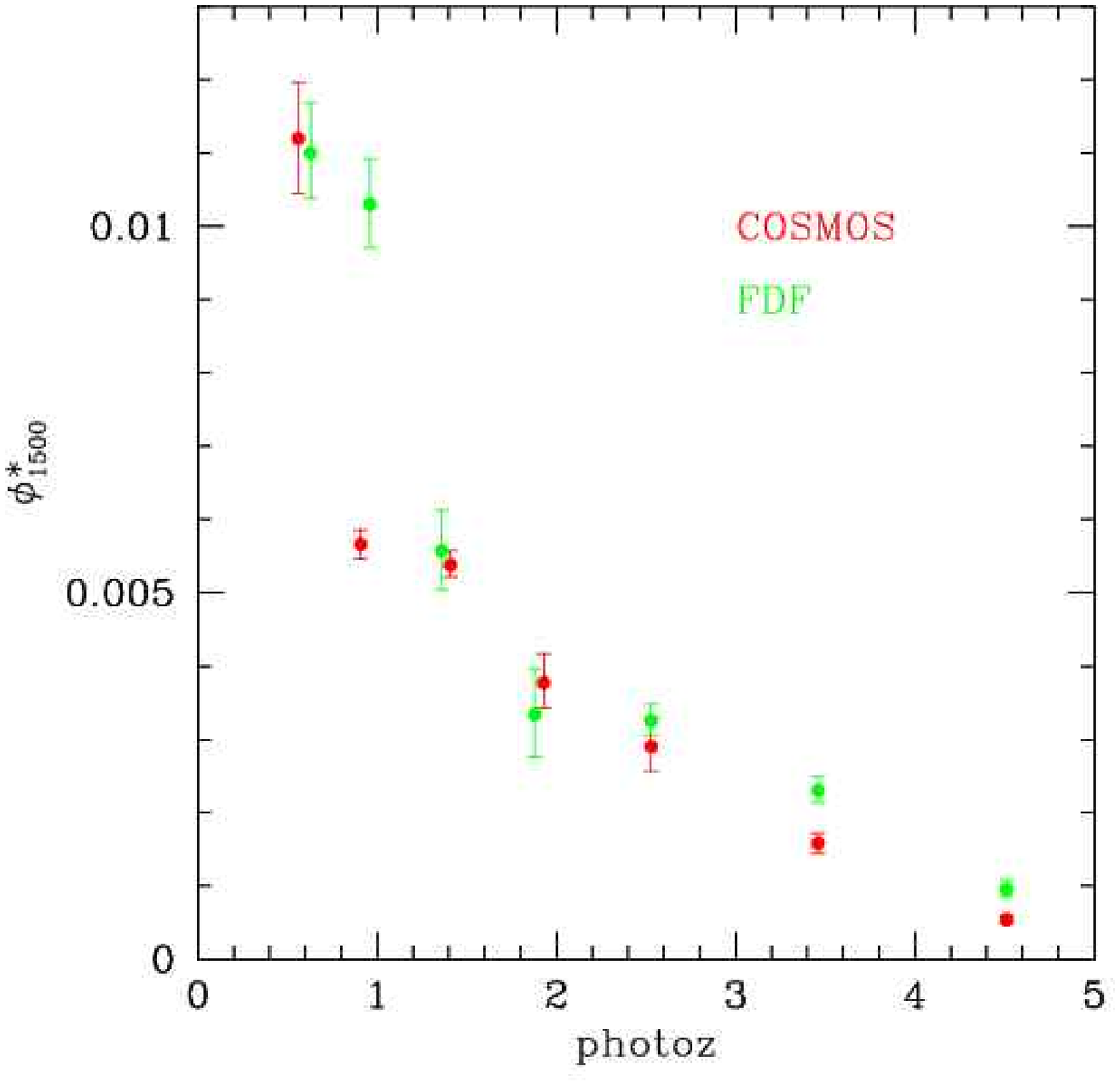}
\caption{Redshift evolution of M$^\ast$ (left panel) and $\phi^\ast$ (right
  panel) for the UV band at 1500~\AA. The red dots are derived from
  the mean luminosity function in the COSMOS field (Schechter fit to
  the red dots in Fig.~\ref{fig:uvlf}) whereas the green dots stem
  from the FDF.
  \label{fig:mstar_phistar_evol}}
\end{figure*}

\begin{table*}
\caption[]{\label{tab:schechter_fit_1500}Schechter function fit at 1500~\AA}
\begin{center}
\begin{tabular}{c|c|c|c}
 redshift interval & M$^\ast$ (mag) & $\phi^\ast$ (Mpc$^{-3}$) & $\alpha$ (fixed)\\
\hline
0.41 -- 0.71 & $-$18.20 +0.09 $-$0.09 & 1.12e-02 +7.61e-04 $-$7.61e-04 &$-$1.07 \\  
0.71 -- 1.11 & $-$19.37 +0.05 $-$0.05 & 5.66e-03 +1.97e-04 $-$1.97e-04 &$-$1.07 \\  
1.11 -- 1.71 & $-$19.66 +0.04 $-$0.04 & 5.38e-03 +1.97e-04 $-$1.69e-04 &$-$1.07 \\  
1.71 -- 2.15 & $-$20.28 +0.08 $-$0.08 & 3.77e-03 +3.94e-04 $-$3.38e-04 &$-$1.07 \\  
2.15 -- 2.91 & $-$20.70 +0.09 $-$0.09 & 2.90e-03 +3.94e-04 $-$3.38e-04 &$-$1.07 \\  
2.91 -- 4.01 & $-$21.01 +0.08 $-$0.09 & 1.58e-03 +1.41e-04 $-$1.41e-04 &$-$1.07 \\  
4.01 -- 5.01 & $-$21.32 +0.14 $-$0.13 & 5.35e-04 +5.63e-05 $-$5.63e-05 &$-$1.07 \\  
\end{tabular}
\end{center}
\end{table*}

In this section we show the UV luminosity function (LF) at 1500~\AA\
as derived from our deep i-selected catalogue and compare it to the
LFs in the FDF \citep{gabasch:1}.  {Note that in this paper we
are not aiming to give a complete analysis of the LF evolution.  A
detailed analysis of the LFs derived in different passbands as well as
the star-formation rate (SFR) together with an elaborate comparison to
the recent literature will be presented in future papers. Here we
limit ourselves to a one-to-one comparison with the FDF luminosity
functions (mostly as a consistency check to our earlier work).  }

To derive the absolute UV band magnitude we use the best fitting SED
as determined by the photometric redshift code. Since the photometric
redshift code works with aperture fluxes, we only need to correct to
total luminosities by applying an object dependent scale factor. For
this scale factor we used the ratio of the I-band aperture flux to the
total flux as provided by SExtractor (MAG\_APER and MAG\_AUTO).  As
the SED fits all observed-frame passbands simultaneously, possible
systematic errors which could be introduced by using K-corrections
applied to a single observed magnitude are reduced \citep[see ][for
more details]{gabasch:1}.

As an example, we plot in Fig.~\ref{fig:absmag_1500} the absolute
UV-band magnitudes against the photometric redshifts of the objects in
the COSMOS patch 06a. Moreover we also show the absolute UV-band
magnitudes as derived in the FDF. Both fields agree very well in their
magnitude distribution, although there are a few relatively bright
objects in the COSMOS patch (about 5 times the area of the FDF) not
seen in the FDF distribution. To check if those objects could be stars
misclassified as galaxies by our star-galaxy separation criterion, we
decided to use a more conservative criterion for separating stars from
galaxies. We changed our criterion from $2\ \chi_{star}^2 <
\chi_{galaxy}^2$ to $\chi_{star}^2 <\chi_{galaxy}^2$ (see above).  As
can be seen in Fig.~\ref{fig:absmag_1500} even this conservative
criterion does not remove a substantial fraction of these relatively
bright objects. Note that because of the larger seeing (0.95\arcsec)
compared to the FDF (0.55\arcsec), there may be more blended objects
in the catalogue.

The luminosity function is computed by dividing the number of galaxies
in each magnitude bin by the volume $V_\mathrm{bin}$ of the redshift
interval.  To account for the fact that some fainter galaxies are not
visible in the whole survey volume, we performed a $V/V_{max}$
\citep{Schmidt1} correction.  The errors of the LFs were calculated by
means of Monte-Carlo simulations and include the photometric redshift
error of every single galaxy, as well as the statistical error
(Poissonian error).  To derive precise Schechter parameters, we
limited our analysis of the LF to the magnitude bin where the
$V/V_{max}$ correction is negligible (red dots in
Fig.~\ref{fig:uvlf}). We also show the uncorrected LF in the various
plots as open circles.  We did not assume any evolution of the
galaxies within the single redshift bins.  The redshift intervals are
approximately the same size in $\ln(1+z)$, and most of the results we
are going to discuss are based on 1000 -- 4000 galaxies per redshift
bin and per patch.

In Fig.~\ref{fig:uvlf} we present the UV luminosity functions at
1500~\AA\ (we evaluate the luminosity function in the rectangular
filter at \mbox{$1500 \pm 100$~\AA}).  The filled (open) symbols
denote the luminosity function with (without) completeness correction
in the different patches.  We also show the $V/V_{max}$ corrected mean
LFs in the COSMOS field as well as the FDF LFs \citep{gabasch:1}.  The
solid red lines show the Schechter function fitted to the luminosity
function (we used a fixed slope of $\alpha=-1.07$ as found in the
FDF). The best fitting Schechter parameter, the redshift binning as
well as the reduced $\chi^2$ are also listed.

\begin{figure*}
\centering
\includegraphics[width=0.49\textwidth]{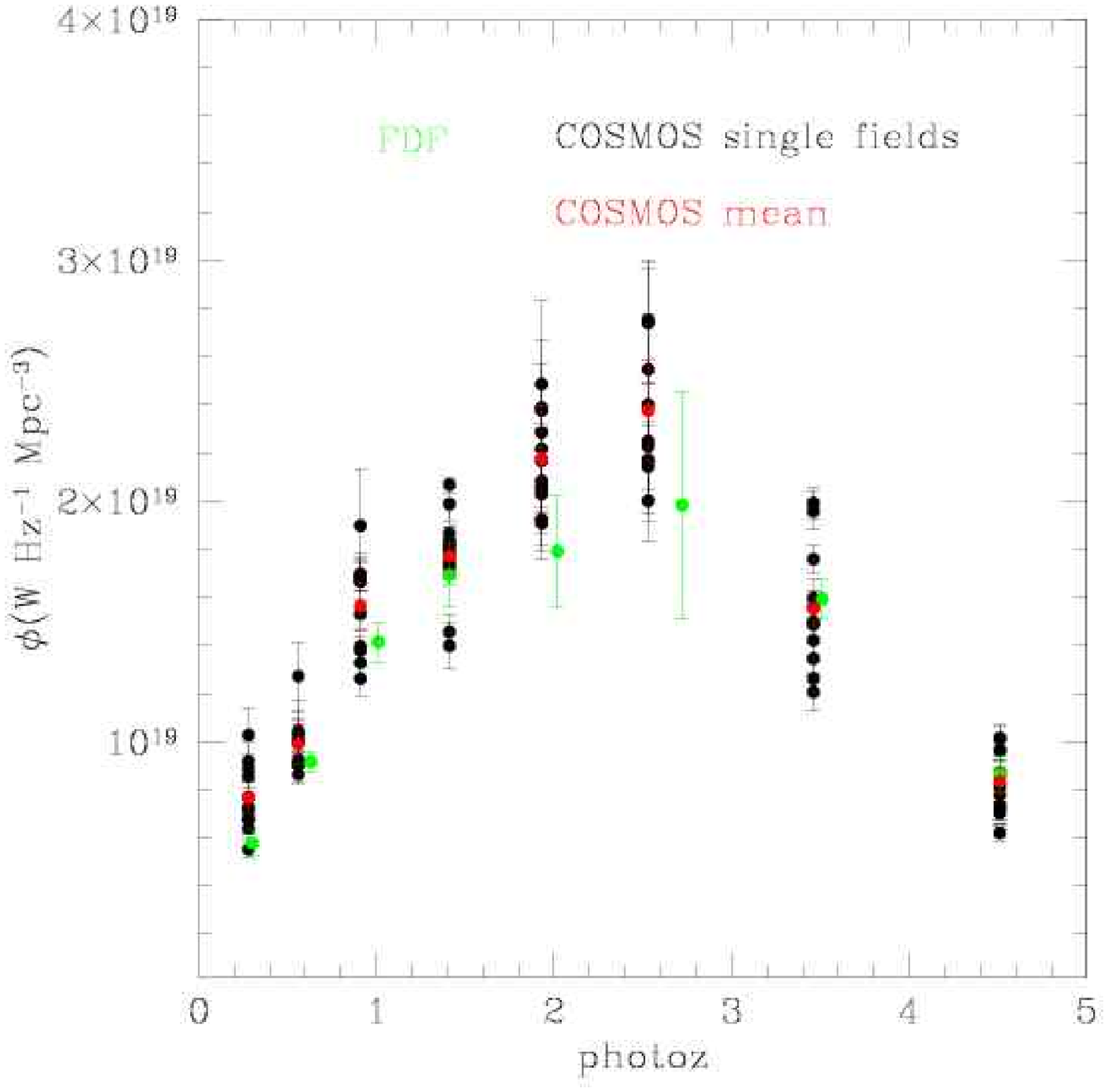} 
\includegraphics[width=0.49\textwidth]{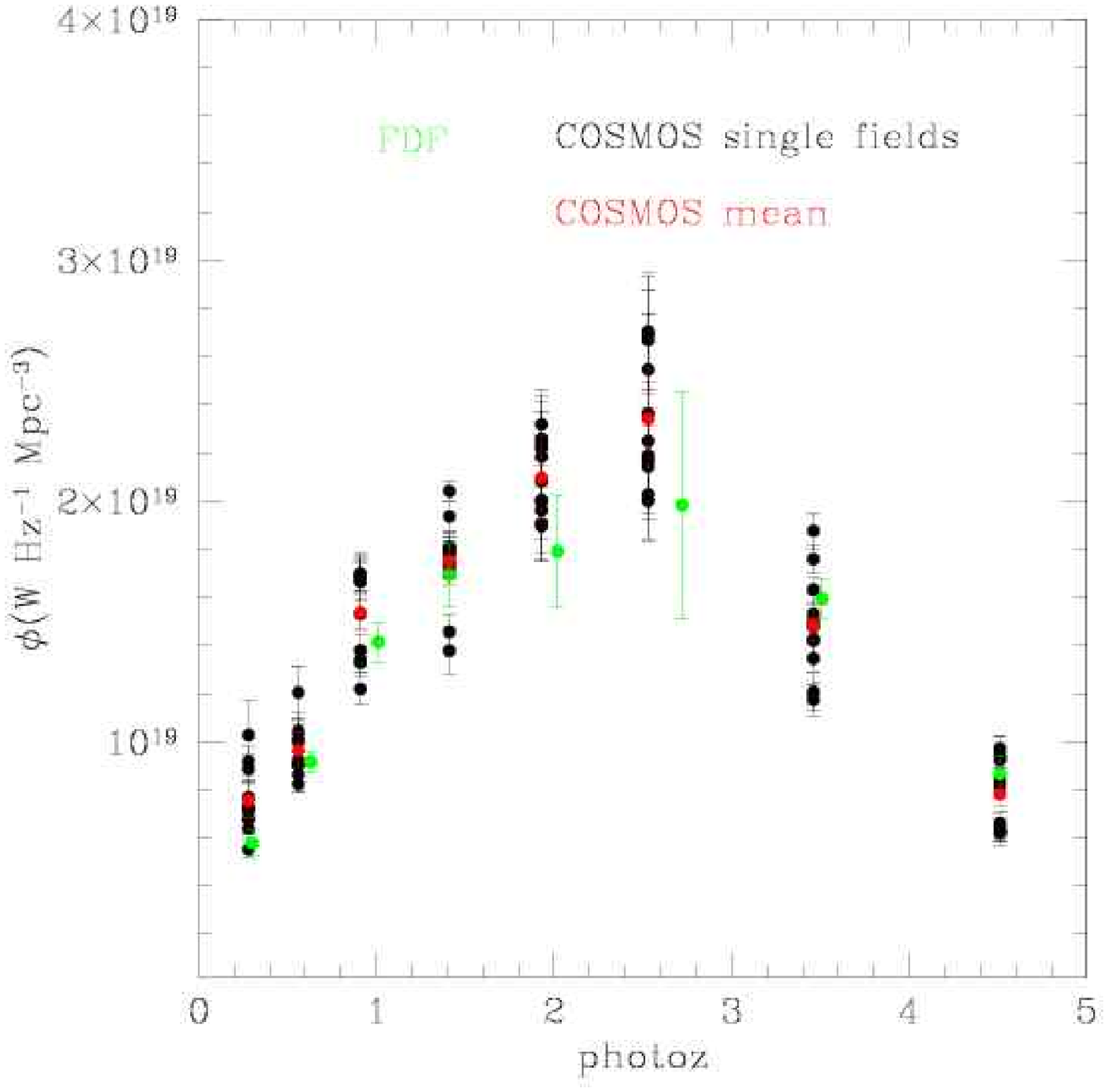} 
\caption{\label{fig:lumdens_1500} 
  The 1500\AA\ luminosity densities of the single COSMOS patches
  (black dots), the mean COSMOS LD (red dots) as well as the LD
  derived in the FDF (green dots) . Left panel: The luminosity
  densities are integrated between the faint-end limiting magnitudes
  given in Table~\ref{tab:lumdens_1500} but \textit{no} limiting
  magnitude cut was used for the bright-end.  Right panel: The
  luminosity densities are integrated between the faint-end and the
  bright-end limiting magnitudes given in Table~\ref{tab:lumdens_1500}.
  The values of the mean COSMOS LD together with the error bars are
  listed in Table~\ref{tab:lumdens_1500}.}
\end{figure*}

It is obvious from the figure that there is strong evolution in
characteristic luminosity and number density between redshifts 0.6 and
4.5. This can be best seen in Fig.~\ref{fig:mstar_phistar_evol}, where
we show the redshift evolution of M$^\ast$ and $\phi^\ast$ as derived
from the Schechter functions fitted to the LFs of Fig.~\ref{fig:uvlf}.
Please note that we used only magnitude bins with a $V/V_{max}$
correction of about unity. Moreover we exclude also very bright
magnitude bins as there might be a contamination by spurious
detections, stars, AGNs or blended objects. The magnitude bins used
for deriving the best fitting Schechter parameters are shown in
Fig.~\ref{fig:uvlf} as red dots. Nevertheless, including also the very
bright bins changes the best fitting values only negligibly.
{We find a substantial brightening of M$^\ast$ and a decrease
  of $\phi^\ast$ with redshift: from \mbox{$\langle z \rangle\sim
    0.5$} to \mbox{$\langle z \rangle\sim 4.5$} the characteristic
  magnitude increases by about 3 magnitudes, whereas the
  characteristic density decreases by about 80 -- 90\%.  Note that our
  results do not disagree with recent findings of e.g.\ 
  \citet{bouwens:2006,bouwens:2007} who are tracing the UV LF from $z
  \sim 3$ to $z \sim 6$ and find a faintening of M$^\ast$ (with
  increasing redshift) and nearly no density evolution in this
  redshift range since we limit our analysis to redshifts of $z \lsim
  4.5$. As discussed in \citet{bouwens:2007} it is plausible that
  there is a turnover in the LF evolution at redshift of $z \sim
  4$.}\\ The best fitting Schechter parameters and their $1\sigma$
errors in the COSMOS field are summarised in
Table~\ref{tab:schechter_fit_1500}. In
Fig.~\ref{fig:mstar_phistar_evol} we also show the redshift evolution
of M$^\ast$ and $\phi^\ast$ as derived in the FDF by
\citet{gabasch:1}. Although in the COSMOS field the characteristic
magnitude is about 0.25 mag brighter in most of the redshift bins if
compared to the FDF results, the characteristic density is slightly
lower.  This results in a very similar UV luminosity density (LD; the
integrated light emitted by the galaxies) in the COSMOS field and in
the FDF, as can be seen in Fig.~\ref{fig:lumdens_1500}.\\ The LD at a
given redshift is derived by summing the completeness-corrected (using
a $V/V_{max}$ correction) luminosity of every single galaxy up to the
absolute magnitude limit.  Contrary to the procedure described in
\citet{gabasch:sfr} and \citet{gabasch:2}, we do \textit{not} apply a
further correction (to zero galaxy luminosity), to take the missing
contribution to the LD of the fainter galaxies into account. This
further correction should be done only after a very careful analysis
of the LFs (as it requires an extrapolation of the LF at the faint
end) and is postponed to a future paper. As the limiting i-band
magnitudes of the COSMOS field and the FDF are very similar, we
decided to integrate the LF in both fields down to the same absolute
magnitude instead (see Table~\ref{tab:lumdens_1500}).  This approach
allows us to directly compare the LD of both fields without relying on
any extrapolation.

In Fig.~\ref{fig:lumdens_1500} we show the 1500\AA\ luminosity
densities of the single COSMOS patches, the mean COSMOS LD as well as
the LD derived in the FDF.  The FDF and COSMOS luminosity densities
are integrated down to the faint-end limiting magnitudes given in
Table~\ref{tab:lumdens_1500}.  Moreover we also show the COSMOS
luminosity densities integrated between the faint-end and a bright-end
limiting magnitude (see Table~\ref{tab:lumdens_1500}). The bright-end
cut excludes in total 80 (0.03\%) bright objects in the COSMOS field.
It was derived by comparing the Schechter-fit to the LF (the excluded
LF bins are shown as blue dots at the bright end in
Fig~\ref{fig:uvlf}) as well as by comparing the absolute UV-band
magnitudes as a function of redshift with those in the FDF (see
Fig.~\ref{fig:absmag_1500} for one patch).
Fig.~\ref{fig:lumdens_1500} nicely shows that the UV LD and its
redshift evolution in the FDF (based on a very small field if compared
to the COSMOS field) agrees quite well with the result in the COSMOS
field. Even though for redshifts below $ z\sim 2.5$ the UV LD in the
FDF is systematically lower than the value in the COSMOS field the
deviation is in the order of $1\sigma$.  The values of the mean COSMOS
LD together with the error bars are listed in
Table~\ref{tab:lumdens_1500}.

\begin{table}
\caption[]{\label{tab:lumdens_1500}
  The mean COSMOS LD at 1500~\AA\ for
  the different redshift bins. The luminosity densities are derived within the
  faint-end and the bright-end limiting magnitudes given in the last column.
}
\begin{center}
\begin{tabular}{c|c|c}
redshift & luminosity density       & magnitude\\
         & (W Hz$^{-1}$ Mpc$^{-3}$) & range\\
\hline
0.15  -- 0.41 &  7.557e+18 $\pm$  7.243e+17 &  -14. --  -22. \\ 
0.41  -- 0.71 &  9.640e+18 $\pm$  7.219e+17 &  -14. --  -22. \\ 
0.71  -- 1.11 &  1.536e+19 $\pm$  7.329e+17 &  -15. --  -23. \\ 
1.11  -- 1.71 &  1.750e+19 $\pm$  6.290e+17 &  -16. --  -23. \\ 
1.71  -- 2.15 &  2.097e+19 $\pm$  1.719e+18 &  -17. --  -23. \\ 
2.15  -- 2.91 &  2.340e+19 $\pm$  2.116e+18 &  -18. --  -24. \\ 
2.91  -- 4.01 &  1.485e+19 $\pm$  6.931e+17 &  -19. --  -24. \\ 
4.01  -- 5.01 &  7.809e+18 $\pm$  4.728e+17 &  -19. --  -24. \\ 
5.01  -- 6.01 &  6.052e+18 $\pm$  8.592e+17 &  -19. --  -25. \\ 
\end{tabular}
\end{center}
\end{table}

\section{Summary and conclusions}
\label{sec:summary_conclusion}

In this paper we present the data acquisition and reduction of NIR Js,
H, and K' bands in the COSMOS field. We describe a 2-pass reduction
pipeline to reduce NIR data. The 2-pass pipeline is optimised to avoid
flat-field errors introduced if the latter are constructed from science
exposures. Moreover we present and implement a method to stack images
of different quality resulting in an optimal S/N ratio for faint sky
dominated {point sources}. The Js and K' band cover an area of
about \mbox{$200\sq\arcmin$} (1 patch) whereas the H band covers about
$0.85\sq\degr$ (15 patches) in total. The 50\% completeness limits are
22.67, $\sim 21.9$, and 21.76 in the Js, H, and K' band, respectively.
The number counts of all NIR bands nicely agree with the number counts
taken from literature.

Furthermore we present a deep and homogeneous i-band selected
multi-waveband catalogue in the COSMOS field by combining publicly
available u, B, V, r, i, z, and K bands with the H band. The clean
catalogue with a formal 50\% completeness limit for point sources of
$i\sim 26.7$ comprises about 290~000 galaxies with information in 8
passbands and covers an area of about $0.7\sq\degr$ (12 patches).  We
exclude all objects with corrupted magnitudes in only one of the
filters from the catalogue in order to have a catalogue as homogeneous
as possible.

Photometric redshifts for all objects are derived and a comparison
with 162 spectroscopic redshifts in the redshift range $ 0 \lsim z
\lsim 3$ shows that the achieved accuracy of the photometric redshifts
is \mbox{$\Delta z / (z_{spec}+1) \lsim 0.035$} with only $\sim 2$\%
outliers. Please note that in order to break the degeneracy between
high redshift and low redshift solutions we included also the GALEX
FUV and NUV filters in the photometric redshift estimation which
considerably reduced the number of outliers.

The multi-waveband catalogue including the photometric redshift
information is made publicly available. The data can be downloaded
from {\verb http://www.mpe.mpg.de/~gabasch/COSMOS/ }

We derive absolute UV magnitudes and a comparison in a
magnitude-redshift diagram with the FDF shows good agreement.
Moreover we investigate the evolution of the luminosity function
evaluated in the restframe UV (1500~\AA). We find a substantial
brightening of M$^\ast$ and a decrease of $\phi^\ast$ with redshift:
from \mbox{$\langle z \rangle\sim 0.5$} to \mbox{$\langle z
  \rangle\sim 4.5$} the characteristic magnitude increases by about 3
magnitudes, whereas the characteristic density decreases by about 80
-- 90\%.

We compare the redshift evolution of the UV luminosity density in the
COSMOS field and the FDF up to a redshift of $z\sim 5$. Below a
redshift of $z\sim 2.5$ the mean UV luminosity density in COSMOS is
systematically higher by about $1\sigma$ if compared to the FDF. At
$2.5 \lsim z \lsim 5$ both UV luminosity densities agree very well.

It is worth noting the remarkably good agreement between the UV LF as
well as the UV LD despite the fact, that the FDF is about 60 times
smaller than the COSMOS field analysed here.

\section*{Acknowledgements} 

{We thank the anonymous referee for his helpful comments which
  improved the presentation of the paper.}  The authors would like to
thank the staff at Calar Alto Observatory for their support during the
observations which were taken in a collaborative effort with the
HIROCS team of Dr. H.-J. R\"oser (MPIA Heidelberg) and the ALHAMBRA
team of Dr. M. Moles (IAA Granada). We thank both teams for the kind
collaboration during the preparation and execution of the program and
Dr. R\"oser and team for useful discussions during data reduction. AG
thanks G.~Feulner, J.~Fliri, and A.~Halkola for a careful and critical
reading of the manuscript and for their valuable suggestions. The
authors thank Ralf Bender for providing the photometric redshift code.
The authors also thanks Nigel Metcalfe for making number count data
available in electronic form.  This work was supported by the
\emph{Deut\-sche For\-schungs\-ge\-mein\-schaft, DFG}, SFB 375
(Astro\-teil\-chen\-phy\-sik).

\label{lastpage}

\end{document}